%% file: arxiv.tex
\documentclass[10pt,twocolumn,letterpaper]{article}

\usepackage[pagenumbers]{cvpr} % To force page numbers, e.g. for an arXiv version

\usepackage{graphicx}
\usepackage{amsmath}
\usepackage{amssymb}
\usepackage{booktabs}
\usepackage{multirow}
\usepackage{enumitem}
\usepackage{xr}
\usepackage{makecell}
\usepackage[pagebackref,breaklinks,colorlinks]{hyperref}

\usepackage[capitalize]{cleveref}
\crefname{section}{Sec.}{Secs.}
\Crefname{section}{Section}{Sections}
\Crefname{table}{Table}{Tables}
\crefname{table}{Tab.}{Tabs.}

\begin{document}

\title{Talking Head Generation with Probabilistic Audio-to-Visual Diffusion Priors}

\author{Zhentao Yu \quad Zixin Yin \quad Deyu Zhou \quad Duomin Wang \quad Finn Wong \quad Baoyuan Wang \\
	{Xiaobing.AI}\\
	{\tt\small \{yuzhentao, yinzixin, zhoudeyu, wangduomin, wangwenlan, wangbaoyuan\}@xiaobing.ai}
}

\twocolumn[{
\renewcommand\twocolumn[1][]{#1}
\maketitle
\begin{center}
    \captionsetup{type=figure}
    \includegraphics[width=1\textwidth]{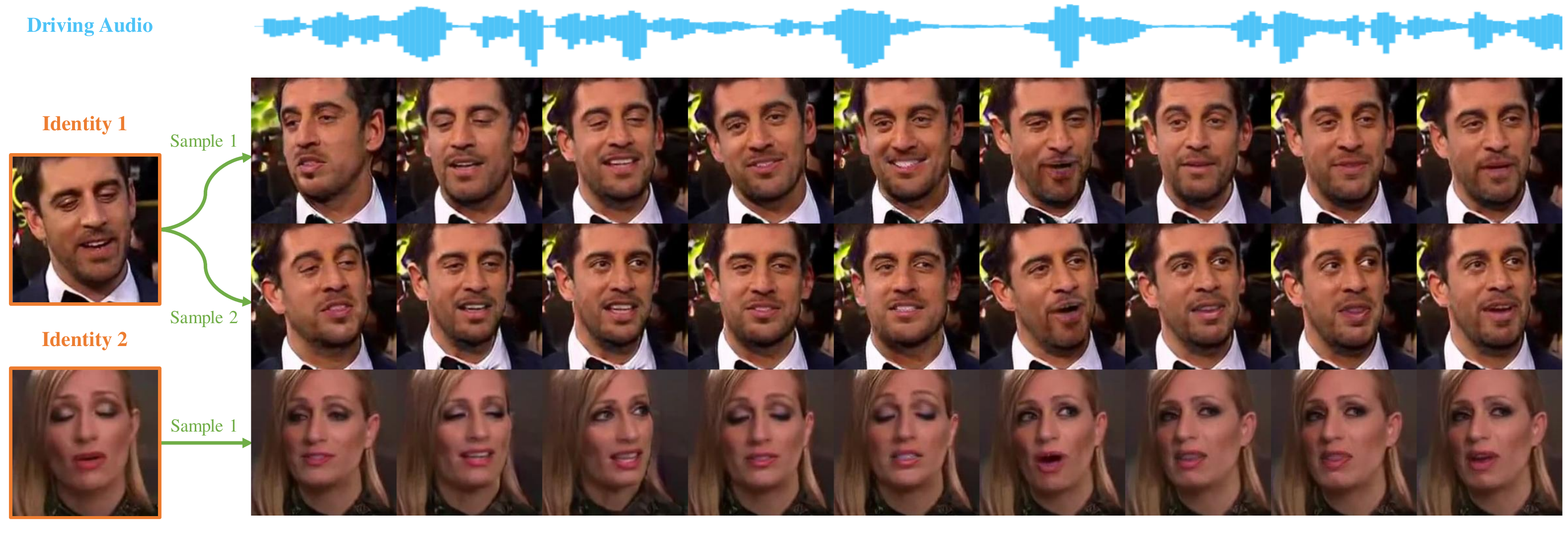}
    \captionof{figure}{Given only an audio source and an arbitrary identity image, our system can generate a video with natural-looking and diverse facial motions (pose, expression, blink \& gaze), while maintaining accurate audio-lip synchronization. Here we show randomly sampled sequences from our diffusion prior for two identities, note that the lip-irrelevant facial motion varies but the lip is still in-sync.}
    \label{fig:teaser}
\end{center}
}]

% \maketitle
% \begin{figure*}[h]
% \centering
% \includegraphics[width=1\textwidth,height=0.256\textwidth]{figures/teaser_v2.pdf}
% \caption{Teaser v2.}
% \label{fig:teaser}
% need to put in the first page.
% \end{figure*}

%%%%%%%%% ABSTRACT
\begin{abstract}
In this paper, we introduce a simple and novel framework for one-shot audio-driven talking head generation. Unlike prior works that require additional driving sources for controlled synthesis in a deterministic manner, we instead probabilistically sample all the holistic lip-irrelevant facial motions (i.e. pose, expression, blink, gaze, etc.) to semantically match the input audio while still maintaining both the photo-realism of audio-lip synchronization and the overall naturalness. This is achieved by our newly proposed audio-to-visual diffusion prior trained on top of the mapping between audio and disentangled non-lip facial representations. Thanks to the probabilistic nature of the diffusion prior, one big advantage of our framework is it can synthesize diverse facial motion sequences given the same audio clip, which is quite user-friendly for many real applications. Through comprehensive evaluations on public benchmarks, we conclude that (1) our diffusion prior outperforms auto-regressive prior significantly on almost all the concerned metrics; (2) our overall system is competitive with prior works in terms of audio-lip synchronization but can effectively sample rich and natural-looking lip-irrelevant facial motions while still semantically harmonized with the audio input. Project page: \url{https://zxyin.github.io/TH-PAD}.
\end{abstract}

%%%%%%%%% BODY TEXT

%-------------------------------------------------------------------------
\input{intro.tex}
\input{relatedwork-v2.tex}

\input{method.tex}
\input{experiment.tex}
\section{Conclusion and Future Work}
In this paper, we introduce a novel talking head generation method based on a diffusion prior model, which can generate diverse and natural-looking talking head videos with only audio and an identity image as inputs. Such ``\textit{audio-driving is all you need}" setting is very friendly for reenactment and dubbing applications. Comprehensive evaluations including our newly proposed metrics validated the effectiveness of our system. As future work, it would be interesting to extend the prior to the whole body for broader audio-driven human reenactment.

%The advantages of our method include 1) Audio driving is all you need. Our model can sample diverse and natural-looking facial motions based on only audio input through an audio2nonlip prior. 2) Accurate lip-audio synchronization. As a possible future work, it would be interesting to extend our method to other generation tasks.
% \begin{conclusion}

% \end{conclusion}
%%%%%%%%% REFERENCES

\clearpage

\twocolumn[{%
\centering
\Large{\textbf{Supplementary Material}}\\[1.5em]
% \Large{\textbf{Supplementary Material}}
}]

\renewcommand\thesection{\Alph{section}}  % make 1 2 3 to A B C
\renewcommand{\thefigure}{\Roman{figure}}
\renewcommand{\thetable}{\Roman{table}}
\renewcommand{\theequation}{\Roman{equation}}
\setcounter{figure}{0}
\setcounter{equation}{0}
\setcounter{section}{0}


\section{Implementation Details}
\subsection{Pre-processing Details}
In the training stage, all input images are cropped and resized to $224\times224$ following the data pre-processing pipeline of~\cite{chung2018voxceleb2} with a face detector~\cite{dlib09} and a landmark detector~\cite{bulat2017far}. For the pre-processing of audio input, we follow~\cite{zhou2021pose} and convert the audio to mel-spectrogram with a sampling rate of 16kHz. Note that for each video frame, we extract an audio segment of 0.2s in the video centered at the video frame to construct an audio-video training sample. 

\subsection{Network Details}
Our main framework consists of 6 modules:
\begin{itemize}[leftmargin=*, itemsep=0pt]
\item \textbf{Identity Encoder \text{E}$_{id}$},  a pretrained ResNeXt50~\cite{xie2017aggregated}.
\item \textbf{Visual Encoder \text{E}$_v$}, also named as non-identity encoder. It is a MobileNetV2~\cite{sandler2018mobilenetv2} following the pretraining scheme of LPD~\cite{Burkov_2020_CVPR}. $\text{MLP}_l$ and $\text{MLP}_{nl}$ are applied after \text{E}$_v$ to map the visual feature $\textbf{f}^v$ into lip space and non-lip space, respectively. The dimension of $\text{MLP}_l$ is $512\times470$ and the dimension of $\text{MLP}_{nl}$ is $512\times42$.
\item \textbf{Audio Encoder \text{E}$_a$}, a ResNet34 encoder from~\cite{chung2020defence}.
\item \textbf{Prior Network \text{P}$_{a2nl}$}, a 6-layer Transformer~\cite{vaswani2017attention} encoder, with 512-d tokens and 1024-d fully forward layers. The positional embeddings are learnable. Note that our diffusion prior and auto-regressive prior networks share the same architecture but with different attention mechanisms, noted as bidirectional attention and causal attention, respectively. The max length of the input tokens of the encoder is set to 128.
\item \textbf{Generator G}, borrowed from StyleGAN2~\cite{karras2020analyzing} and has the same modulated convolution as mentioned in~\cite{zhou2021pose}.
\item \textbf{Discriminator D}, same as the discriminator used in~\cite{zhou2021pose}.
\end{itemize}

Three other pre-trained models are utilized during reconstruction learning and quantitative evaluation, including:
\begin{itemize}[leftmargin=*, itemsep=0pt]
\item \textbf{Gaze Encoder~\cite{abdelrahman2022l2cs}}, the last 512-d feature of the encoder is used for the calculation of $L_{gaze}$.
\item \textbf{VGG Network~\cite{simonyan2014very} }, for the calculation of VGG loss in~\cite{zhou2021pose}.
\item \textbf{Deep3DFace~\cite{deng2019accurate}}, a 3DMM model extracting the 3-d  pose and 64-d expression coefficients for the evaluation of $\textbf{FID}_{fm}$, $\textbf{FID}_{{\Delta}fm}$ and $\textbf{SND}$.
\end{itemize}

% In the training phase, a batch consists of 1 positive and 8 negative data pairs, where each pair represents one video frame and the corresponding 0.2s audio segment.
\subsection{Loss Details}
\paragraph{Lip \& Non-lip Disentanglement}
\begin{itemize}
\item \textbf{Audio-Visual Contrastive Learning}
We use the same implementation as in CLIP~\cite{radford2021learning} for the contrastive learning of audio encoder \text{E}$_{a}$ and a pre-trained visual encoder \text{E}$_{v}$~\cite{Burkov_2020_CVPR}.  We utilize the audio and frames from the same video to construct a contrastive batch, where the corresponding pairs are positive pairs. Our models were trained on 4 A100 GPUs for 30 epochs with batch size of 288. The initial learning rate is set to $1e^{-5}$ with a decay rate of 0.93 for every 200,000 steps.
\item \textbf{Reconstruction Learning for Non-Lip Space} The loss formulas are listed as follows~\cite{Burkov_2020_CVPR}~\cite{zhou2021pose}:
\begin{flalign}
    \begin{split}
    L_{\text{GAN}}  =& ~\min_{\text{G}}\max_{\text{D}}\sum_{n=1}^{N_{D}}\mathbb{E}_{I_{(i)}}[\log{\text{D}_{n}}({I_{(i)}})] \\
              & +\mathbb{E}_{\textbf{f}_{cat(i)}}[\log({1-\text{D}_{n}}(\text{G}(\textbf{f}_{cat(i)})))],
    \end{split} \\
    \begin{split}
    L_{L1}   =& ~\sum_{n=1}^{N_{D}}\left\|\text{D}_{n}(I_{(i)})-\text{D}_{n}(\text{G}(\textbf{f}_{cat(i)}))\right\|_1,
    \end{split}\\
    \begin{split}
    L_{\text{VGG}}  =& ~\sum_{n=1}^{N_{G}}\left\|\text{VGG}_{n}(I_{(i)})-\text{VGG}_{n}(\text{G}(\textbf{f}_{cat(i)}))\right\|_1,
    \end{split} \\
    \begin{split}
    L_{G}   =& ~\lambda_{ol}\cdot L_{ol} + \lambda_{gaze}\cdot L_{gaze} + \lambda_{L1}\cdot L_{L1} \\
              & +\lambda_{\text{GAN}}\cdot L_{\text{GAN}} + \lambda_{\text{VGG}}\cdot L_{\text{VGG}},
    \end{split}
\end{flalign}
where all $\lambda$s are set to 1. $I_{(i)}$ and $\text{G}(\textbf{f}_{cat(i)})$ are the GT image and the generated image of  the $i$-th sample, respectively. $L_{\text{GAN}}$ is a multi-scale generative adversarial loss. $N_{D}$ is the number of layers in discriminator D and the subscript $n$ implies the $n$-th layer of D. $L_{L1}$ is the L1 distance between the $n$-th layer features of GT and generated image extracted from D. Similar to $L_{L1}$, $L_\text{VGG}$ is defined as the L1 distance between two features extracted from a pre-trained VGG network. $L_{gaze}$ and $L_{ol}$ are described in the paper. 

In the disentanglement of lip and non-lip, the learning rate of two MLPs of $\text{E}_{v}$ is initialized as $1e^{-5}$ with a decay rate as 0.5 for every 80,000 steps. The learning rates of G and D are initialized as $2e^{-5}$ and $3.5e^{-6}$, respectively, with the same decay rate. The batch size is set to 16 and the size of \textbf{MB}, $K$, is set to 32 which stores 512 features in total. After the 40k-th step, we freeze other modules and train only G and D for 5 epochs on 4 A100 GPUs.

\end{itemize}

\paragraph{Audio2Visual Prior}
After the disentanglement of lip \& non-lip, we trained the prior network \text{P}$_{a2nl}$ along with the pre-trained visual encoder $\text{E}_{v}$ and audio encoder $\text{E}_{a}$, where $\text{E}_{v}$ extracts the non-lip feature $\textbf{f}_{nl}^v$ and $\text{E}_{a}$ extracts the audio feature $\textbf{f}_{l}^a$. 

\begin{itemize}
    \item \textbf{Diffusion Prior}
        The input of $\text{P}_{a2nl}$ is a concatenated feature $cat(a_{1:L},n_{1:L}^t)$, where $a_{1:L}$ is an audio feature sequence and $n_{1:L}^t$ is a non-lip feature sequence with random noise added through the diffusion process. $t$ is the time step and $L$ is the length of the sequence. During training, $t$ is uniformly sampled in [1, T] where $T$ is set to 1000. MSE loss is calculated between the output of  $\text{P}_{a2nl}$ and the GT non-lip feature. At inference time, $n_{1:L}^t$ is replaced with random Gaussian noise. Note that bidirectional attention is used in $\text{P}_{a2nl}$, implying that denoised features are outputted at the same time.  While trained with mask editing, $n_{1:L}^t$ is randomly replaced with GT values such that $\text{P}_{a2nl}$ is trained to predict the unmasked region. As a result, the model is capable of interpolating non-lip sequences according to input audio and unmasked non-lip features, or generating non-lip sequences with input audio only at inference time. The model was trained on 1 A100 GPU for 50,000 steps with the batch size as 64 and $L$ as 128. The learning rate is set to $1e^{-4}$.
    \item \textbf{Auto-Regressive (AR) Prior}
        The prior $\text{P}_{a2nl}(\check{n}_{i}|\{\check{n}_{k<i}\}, a_{1:L})$ is modeled in an auto-regressive way where $k=[1,2, ...,i-1]$ and $i=[1, 2, ..., L]$. Different from the diffusion prior, AR prior encodes the input with causal attention instead of bi-directional attention such that tokens can only refer to the previous tokens but not the succeeding ones. MSE loss is used to measure the prediction error of GT and the prediction. We use the same training setting as diffusion prior, \ie batch size, etc. 
        %The main difference between AR prior and the diffusion prior is that the diffusion prior is capable of one-to-many mapping while AR prior is deterministic. 
\end{itemize}
\begin{figure}[!t]
\centering
\includegraphics[width=0.49\textwidth,]{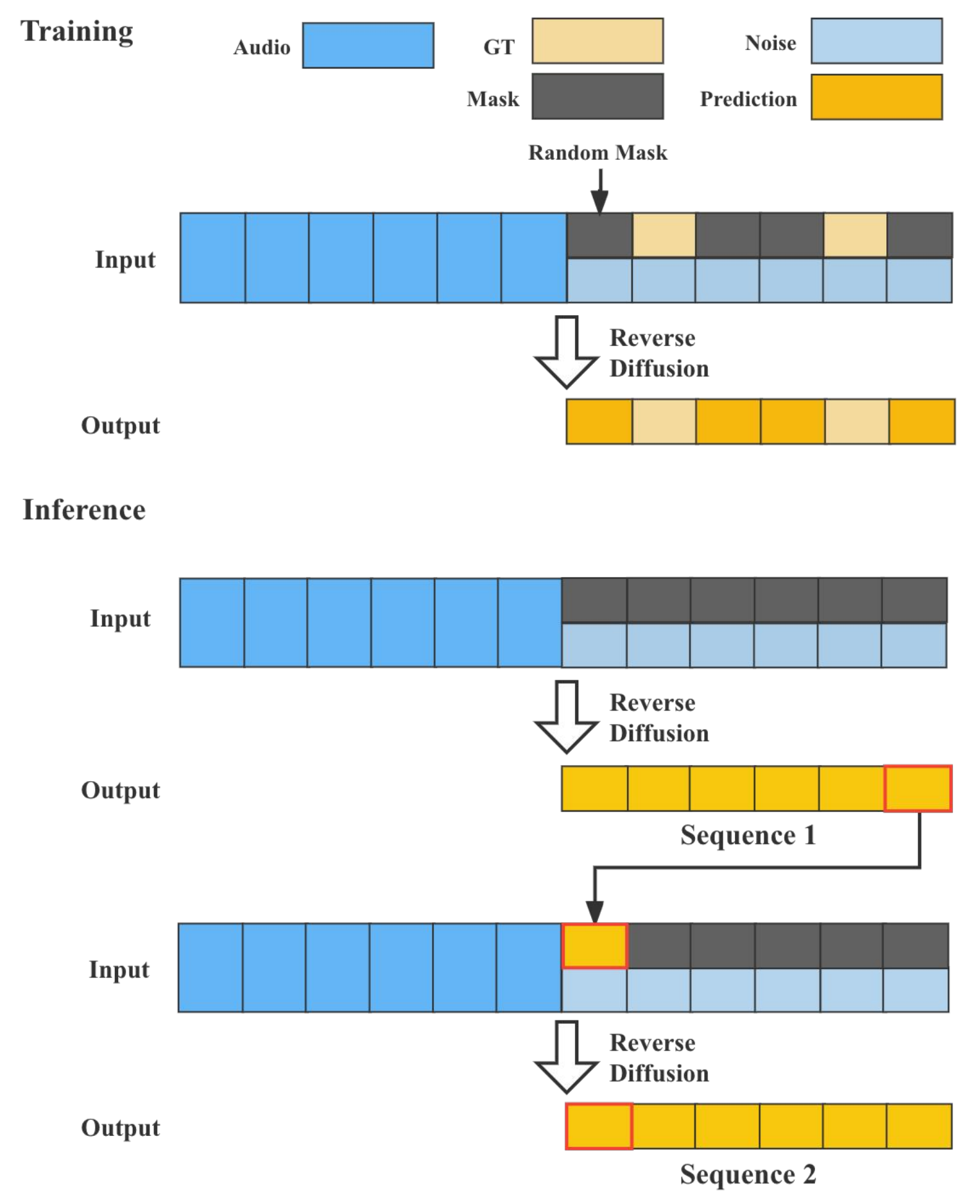}
\caption{Conceptual illustration of our mask editing technique during both training and inference. Here, ``GT" denotes the non-lip feature from $\text{E}_v$. During training, we randomly masked 90\% frames of the non-lip features and conditioned on the rest and audio input as well as noise to predict the masked non-lip features through the reverse diffusion process. During the inference, we feed the predicted non-lip feature of the last frame from the previous sequence as the unmasked non-lip features of the first frame in the next sequence, conditioned on which the subsequent masked non-lip features will be predicted, therefore smooth transition is achieved between two consecutive sequences. }
\label{fig:maskediting}
\vspace{-2mm}
\end{figure}

\vspace{-2mm}

\subsection{Details about Mask Editing Mechanism}
Figure~\ref{fig:maskediting} illustrates how our mask editing works during both the training and inference. As shown, our mask mechanism is very similar to the mask language modeling that is commonly used in Bert ~\cite{devlin2018bert} and other pre-trained models, such as BeiTs ~\cite{BEITV1,bevt2,Beitv3}. Intuitively, we want to primarily count on the audio to infer the non-lip features, however, due to the weak correlation between audio and non-lip facial motions, additional conditions need to feed into the diffusion model to reduce the ambiguities of the mappings. As illustrated, during training, we empirically copy 10\% frames of non-lip features from GT and ask the model to predict the rest 90\% masked frames. This is inspired by the image inpainting works \cite{rombach2022high}. We leave a thorough study of the masking mechanism as future work.

\subsection{Evaluation Details}
\paragraph{Baselines}
Only methods that have pre-trained models released were chosen as baselines for fair comparisons. These methods are introduced as follows,
\begin{itemize}[leftmargin=*,itemsep=-2pt]
\item \textbf{Wav2Lip~\cite{wav2lip}} generates the lower-half face given an identity image, an upper-half driving image, and an audio clip. Other facial regions remain unchanged.\\
\vspace{-4mm}
\item \textbf{MakeItTalk~\cite{zhou2020makelttalk}} learns an identity-specific embedding and a speech-content embedding to predict facial landmarks. Face warping and image translation are applied afterward for face reenactment.\\
\vspace{-4mm}
\item \textbf{PC-AVS~\cite{zhou2021pose}} takes an identity image, a reference pose video, and an audio clip as inputs to generate a talking head video. It does not support other facial motions such as expression and eye blinks.\\
\vspace{-4mm}
\item \textbf{Audio2head~\cite{wang2021audio2head}} learns to predict the head pose auto-regressively given a reference image and an audio clip and then generates an audio-driven talking head accordingly.\\
\vspace{-4mm}
\item \textbf{EAMM~\cite{ji2022eamm}} generates a talking head video from an emotion video, a driving audio, an identity image, and a pose sequence. 
\end{itemize}

\begin{figure}[h]
\centering
\includegraphics[width=0.48\textwidth,]{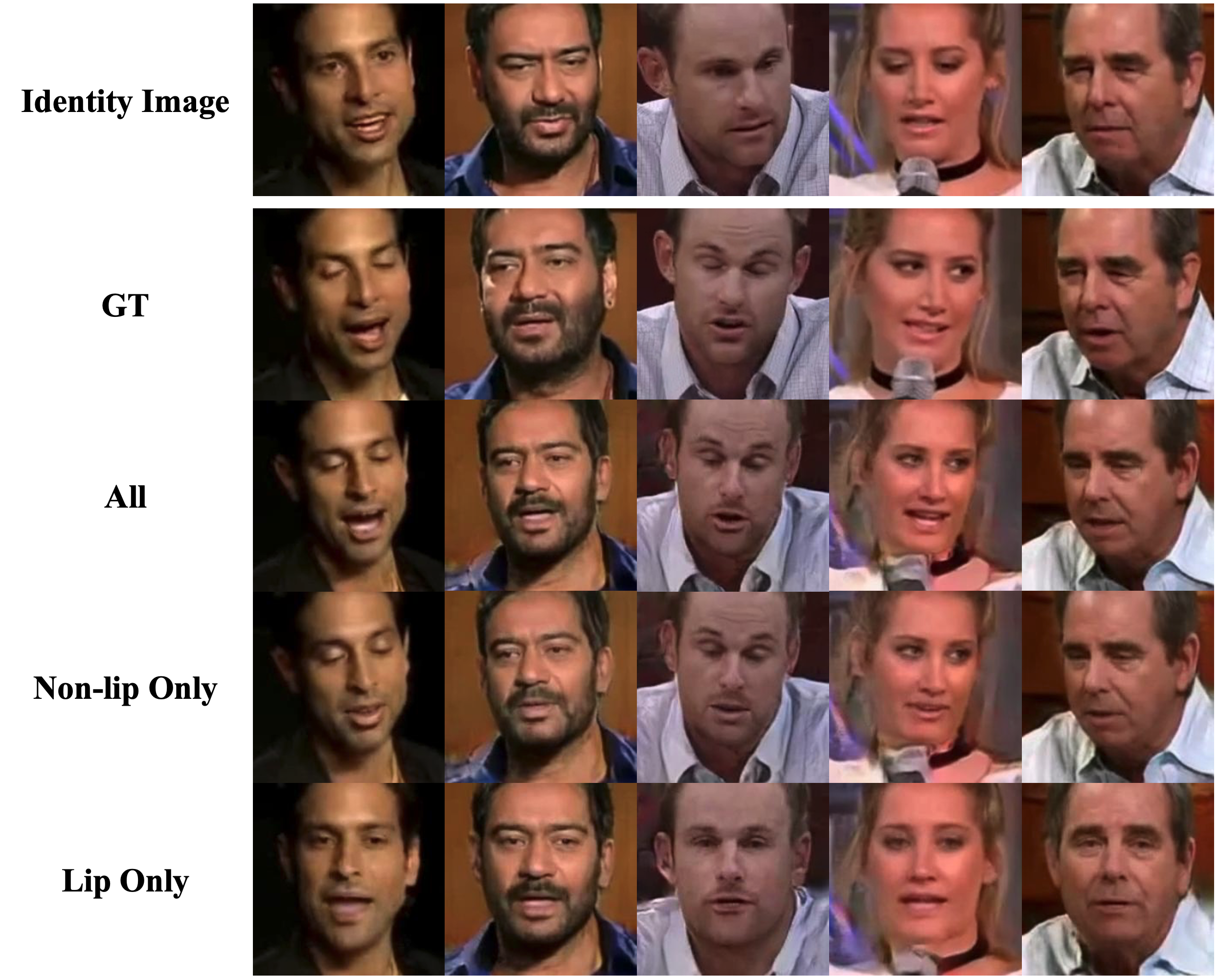}
% \caption{Qualitative results on VoxCeleb2~{chung2018voxceleb2} for non-lip signal only, lip signal only and both in self-reenactment scenario. Note the facial motions of blinks, lip, gaze and expression as represented with yellow, blue, red and green arrows, respectively.}
\caption{Qualitative results on VoxCeleb2~\cite{chung2018voxceleb2} for non-lip signal only, lip signal only, and both of them, respectively.}
\label{fig:disentangle}
\end{figure}

\begin{figure}[h]
\centering
\includegraphics[width=0.48\textwidth,]{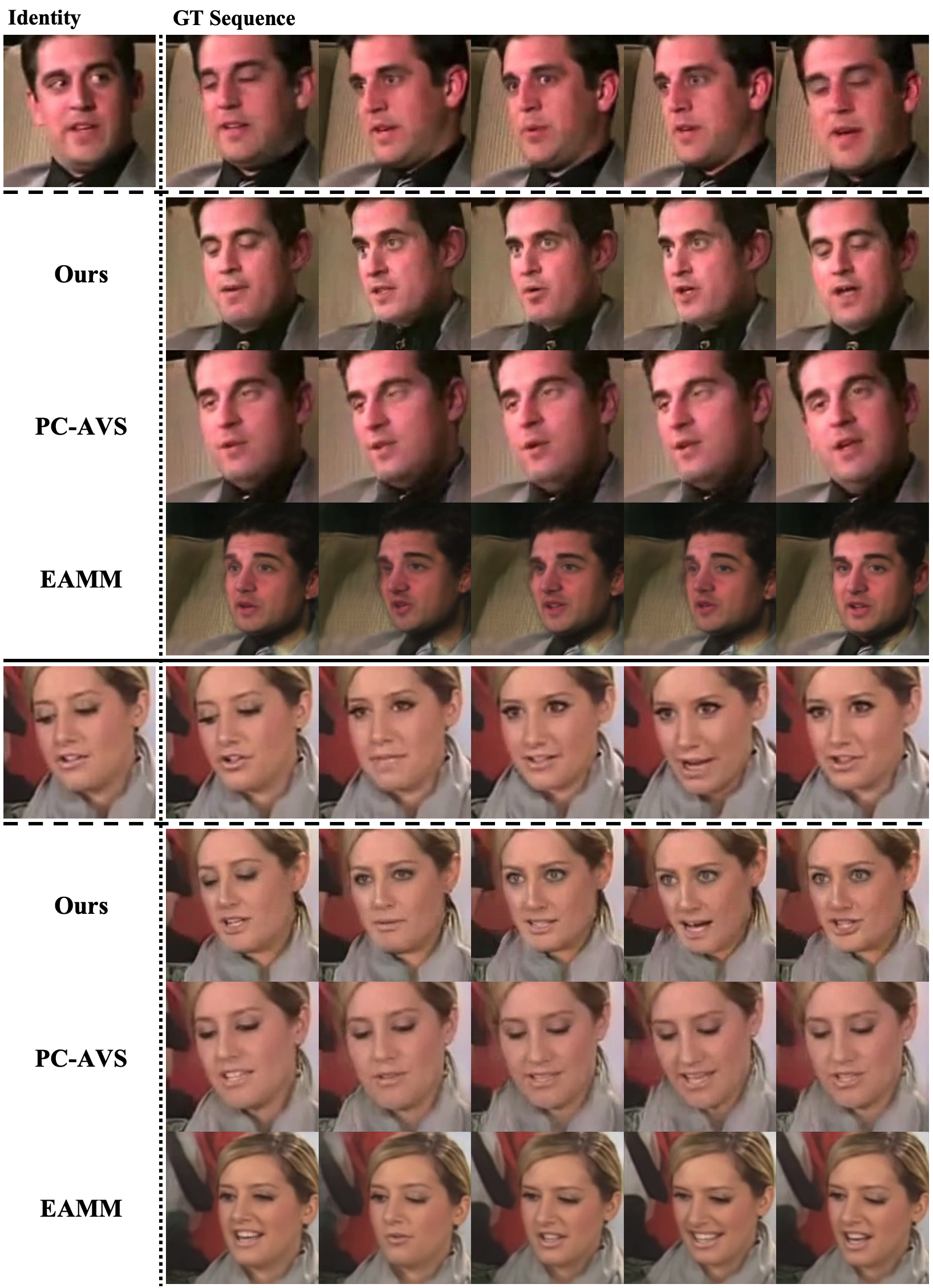}
\caption{Qualitative results on VoxCeleb2 with driving signals from the same video-audio pair as the identity image. Note that our model uses non-lip signals from video input, instead of the diffusion prior. Each row shows five uniformly sampled frames from videos.}
\label{fig:reenact}
\end{figure}

\begin{figure}[h]
\centering
\includegraphics[width=0.48\textwidth,]{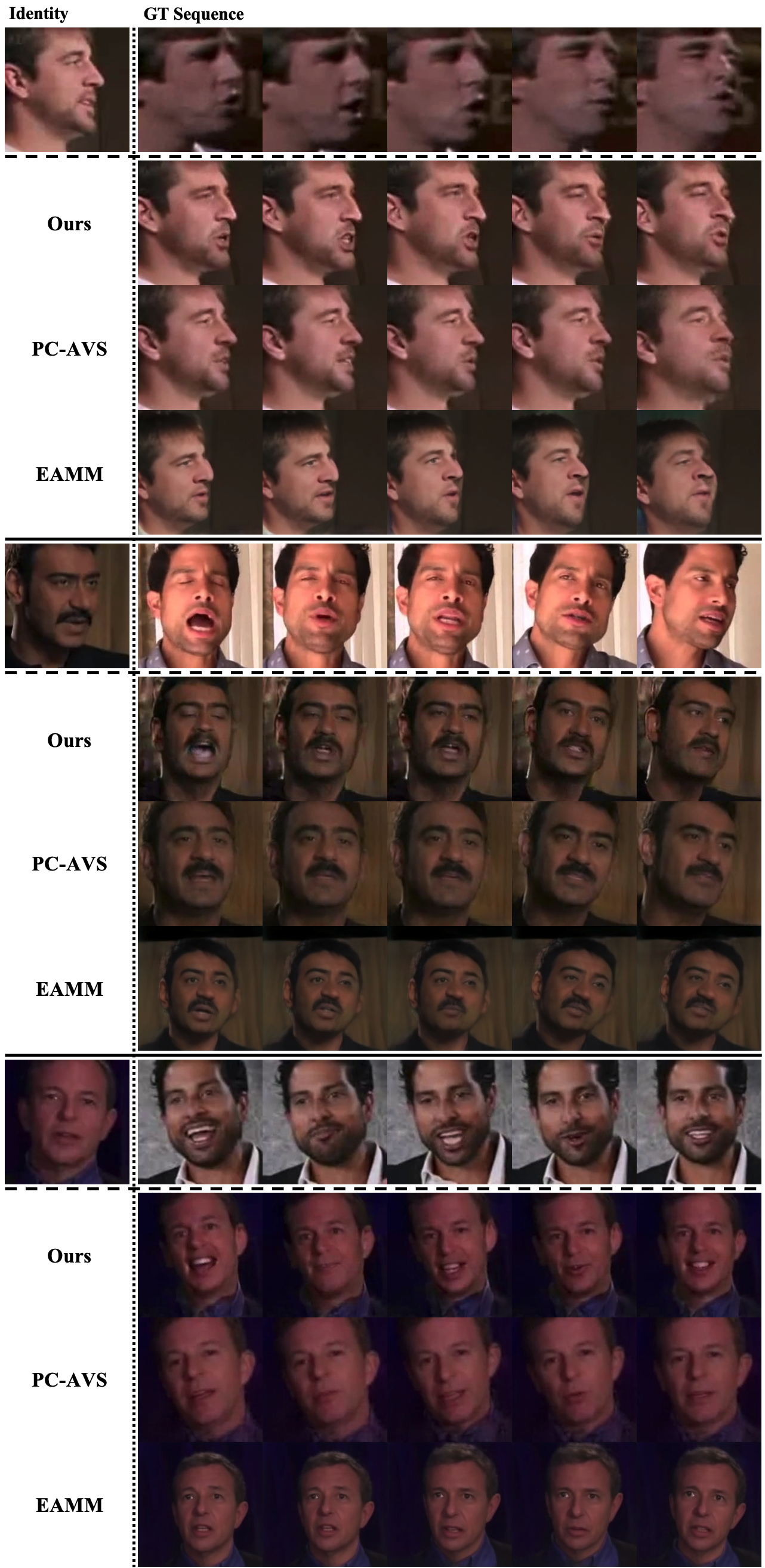}
\caption{Qualitative results on VoxCeleb2 with driving signals from another video-audio pair. Note that our model uses non-lip signals from video input, instead of the diffusion prior. Each row shows five uniformly sampled frames from videos.}
\label{fig:cross}
\end{figure}

\begin{figure}[h]
\centering
\includegraphics[width=0.5\textwidth,]{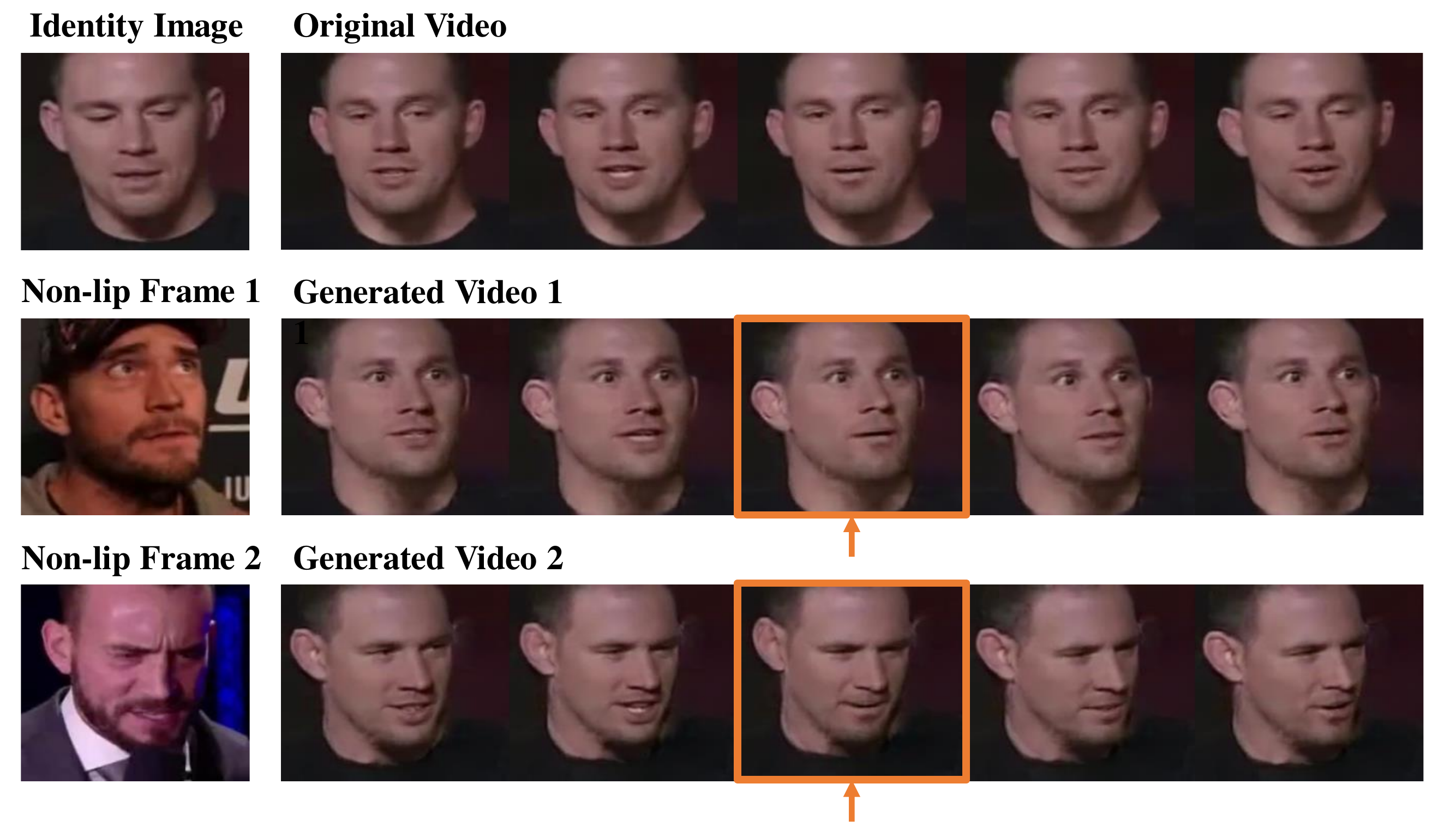}
\caption{Qualitative results of video editing. Each row shows five continuous frames in each video, where the frame with an orange box serves as the conditional non-lip features copied from the example frames shown on the left. Our model predicts smoothly transited sounding frames.}
\label{fig:interpolation}
\end{figure}

\begin{figure}[h]
\centering
\includegraphics[width=0.48\textwidth,]{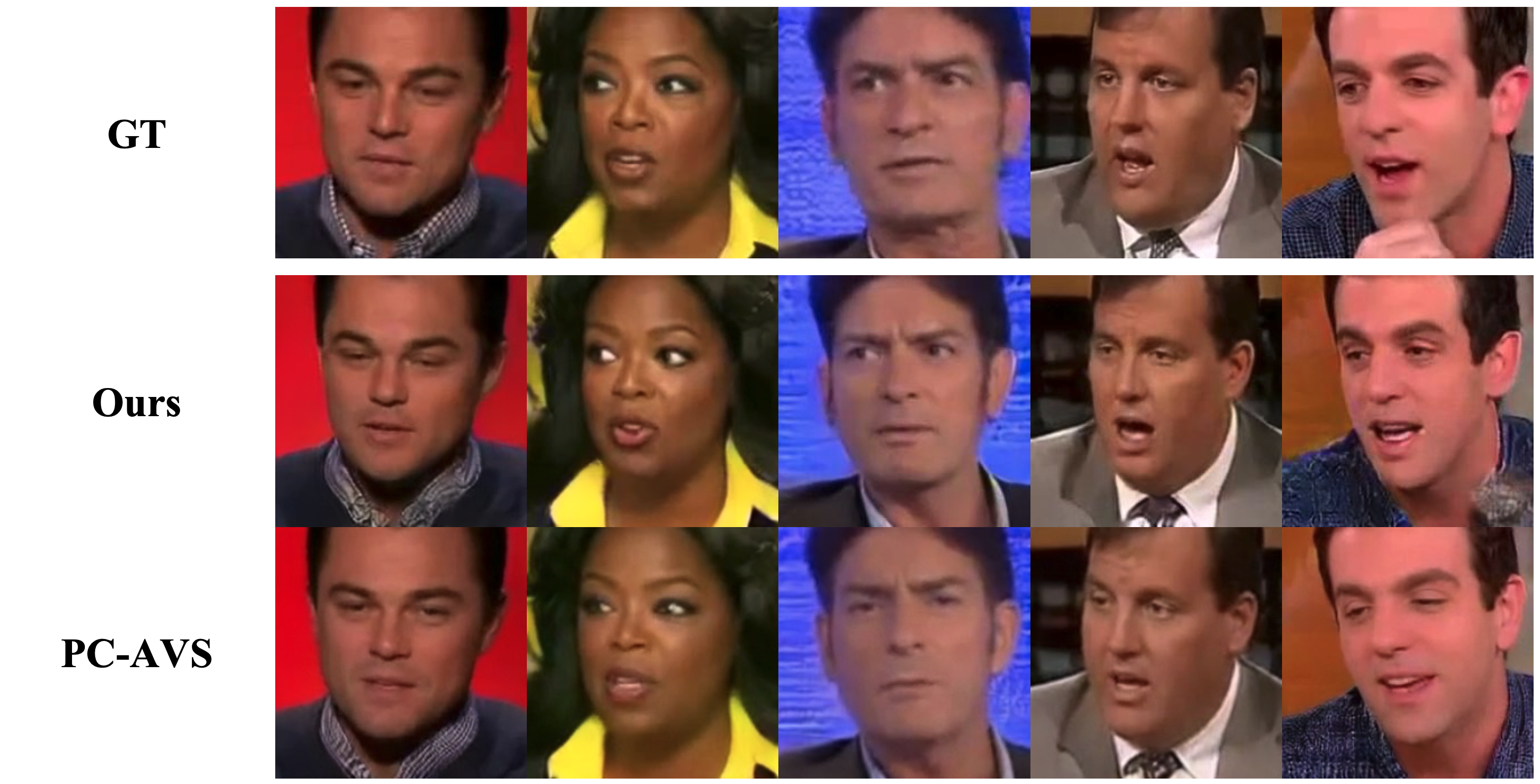}
\caption{Rendering cases compared with PC-AVS on VoxCeleb2 with driving signals from the same video-audio pair as the identity image. Note that our model uses non-lip signals from video input, instead of the diffusion prior.}
\label{fig:short}
\end{figure}

\begin{figure}[h]
\centering
\includegraphics[width=0.5\textwidth,]{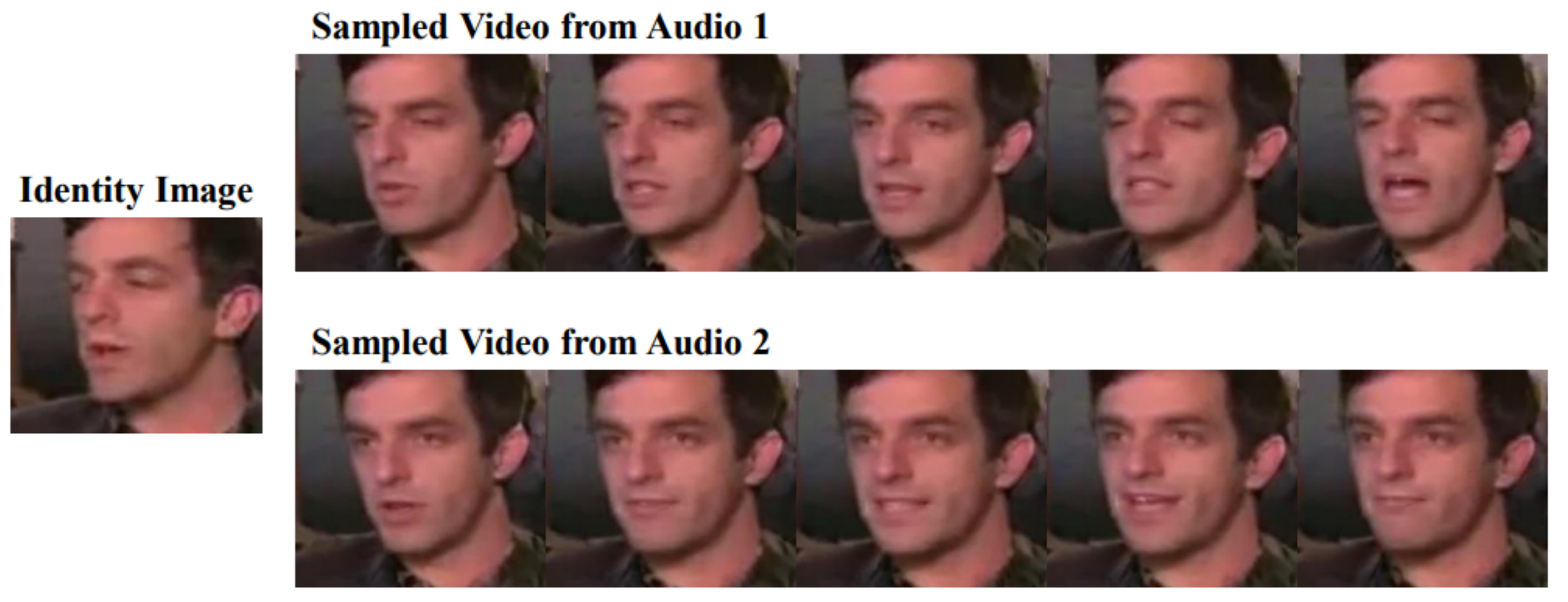}
\caption{Qualitative results of generating videos with the different audios. Each row shows five continuous frames in each video.}
\label{fig:audio}
\end{figure}
% These methods are selected as baselines: \textbf{Wav2Lip~\cite{wav2lip}} produces the lower-half face given an identity image and an audio clip. \textbf{MakeItTalk~\cite{zhou2020makelttalk}} learns to predict facial landmarks and generate head and mouth movement. \textbf{PC-AVS~\cite{zhou2021pose}} can generate a pose-controllable talking face with audio and head pose driving. \textbf{Audio2head~\cite{wang2021audio2head}} learns to predict the head pose autoregressively given a reference image and an audio clip and generates the audio-driven talking head accordingly. \textbf{EAMM~\cite{ji2022eamm}} generates a talking head video from an emotion video, a driving audio, an identity image, and a pose sequence. Other not selected works either because their models are not released or additional driving sources are required.

\paragraph{Driving Settings}

Self-reenactment means driving an identity image with all signals from the same video clip of GT. Different from self-reenactment, cross-reenactment uses another video clip of GT as non-lip signals.

\paragraph{Lip \& Non-Lip Disentanglement}

To evaluate the disentanglement between lip and non-lip on VoxCeleb2, we measure the normalized distance between the 2D landmarks~\cite{yu2021heatmap} of GT images and generated images, for blink and gaze, respectively.

\section{Analysis and Results}
\subsection{Lip \& Non-lip Disentanglement}

Fig.~\ref{fig:disentangle} shows that non-lip signals can drive pose, expression, blink and gaze well with the mouth slightly opened. On the other hand, lip signals can only drive lip motions with others fixed.

We compare our proposed method with others in talking head reenactment, resulting in Fig.~\ref{fig:reenact}. It can be observed that our method can control non-lip motions including pose, expression, blink, and gaze while lip motion is in-sync with the audio. It indicates that our method benefits from a well-disentangled motion space, which is a good foundation for our one-to-many diffusion prior.

Besides, we also showcase cross-id compared with other baselines as shown in Fig.~\ref{fig:cross}. Our proposed method has more diverse motion than others and can control pose, expression, blink and gaze well.

\begin{figure*}[h]
\centering
\includegraphics[width=1\textwidth]{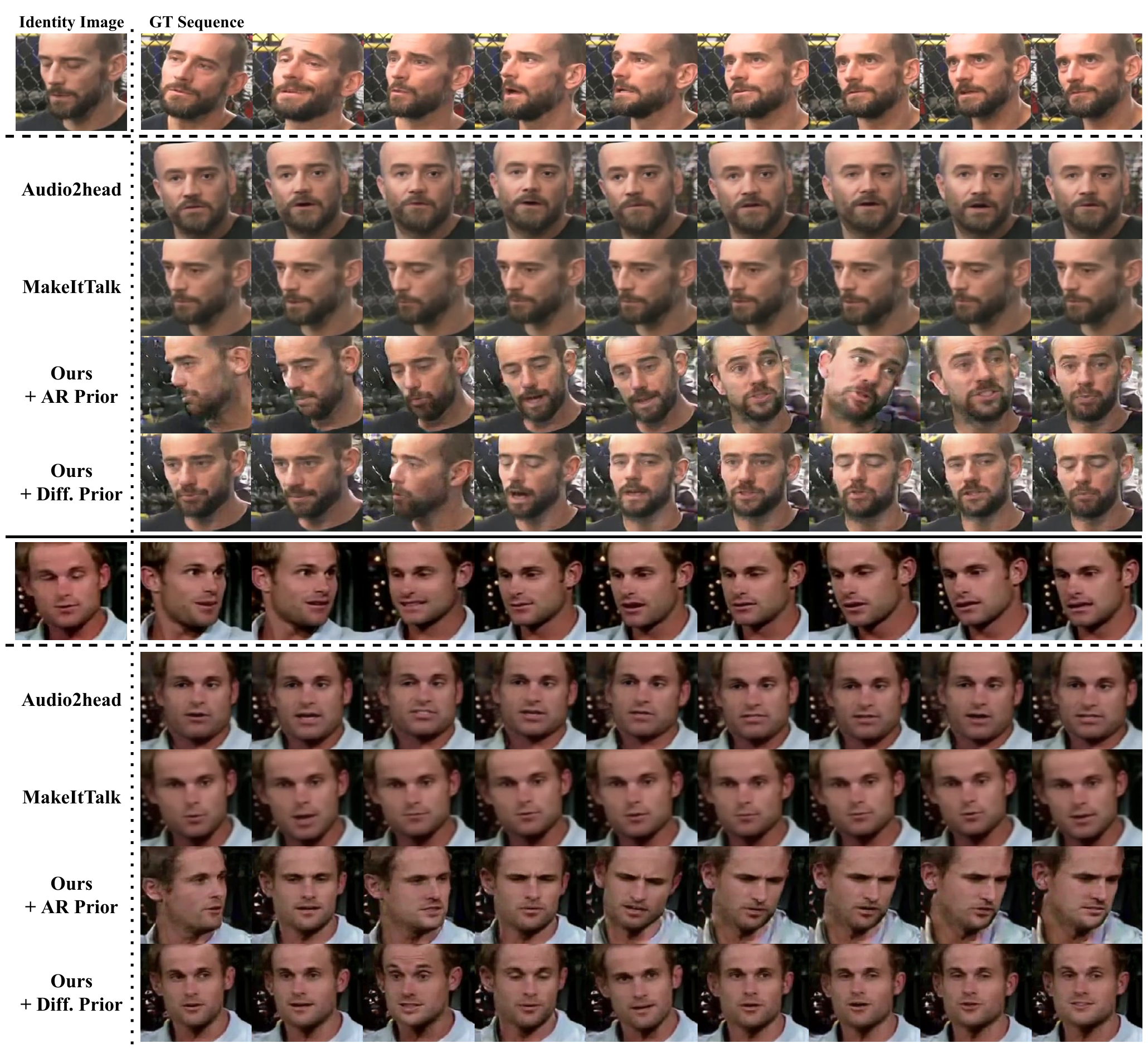}
\caption{Qualitative results of our method as compared to other baselines. Each row shows nine uniformly sampled video frames. Here we use two different audio sources to drive the identities. Our method shows accurate lip-audio synchronization with diverse and natural facial motions.}
\label{fig:gallery2}
\end{figure*}

\subsection{Video Editing Conditioned on Desired Non-lip Feature}
Fig.~\ref{fig:interpolation} shows that our diffusion prior $\text{P}_{a2nl}$ trained with mask editing technique can enable controlled video editing with conditional non-lip features. \ie we can fix the non-lip feature of one particular frame while letting the diffusion prior model predict the non-lip features of the rest frames. In this example, we borrow two non-lip features (extracted by $\text{E}_{v}$) from a randomly chosen frame of different identities. Then, we condition the diffusion prior with these non-lip features and assign them to a particular frame (the 3rd frame in this example, marked with an orange box) to let the model predict the rest frames. As we can see, both resulting sequences (2nd and 3rd rows) respect the conditional input and perform smooth transitions surrounding the conditional frame. This indicates both the robustness and flexibility of our system.

%The first row shows five sequential frames of a video and we extract one frame in the video as the identity image. The other two rows show another two identity images and corresponding non-lip feature sequence extracted by $\text{E}_{v}$, we use interpolate the third frame in the sequence, we randomly choose one frame from other videos with different identities, and use its non-lip feature extracted by $\text{E}_{v}$ as input of model $\text{P}_{a2nl}$. Thus, it generates other non-lip features by the reversed diffusion process. Both the second and third rows show a smooth interpolated video. 

\subsection{Limitations of Rendering}
Although we trained the generator G similarly to that in PC-AVS~\cite{zhou2021pose}, our \textbf{FID} is slightly higher. As shown in Fig.~\ref{fig:short}, our generator may produce stripe artifacts in regions with high-frequency details, {i.e.} clothes and background, thus reducing \textbf{FID}. This is likely caused by our rendered images having a larger proportion on the face than PC-AVS's. Another possible reason is the imperfect disentanglement within the pipeline. Nevertheless, our major focus is to predict the non-lip facial motions based on the audio sequence with the help of a probabilistic diffusion prior model. We leave the improvement of rendering to future works.
\subsection{More Results}
Fig.~\ref{fig:audio} shows the generated videos sampled by different audios with the same identity image. Fig.~\ref{fig:gallery2} shows more comparing results of our model and other baselines. Please refer to our supplementary video for more visual results. 
% \subsection{Curve of Sampled Expression and Pose}
% TBD.

% In addition, the example in the second column indicates that the sharpness of our generated image is higher than that of PC-AVS, which will highlight the abnormalities in dense textures, such as hair.

\clearpage
\appendix

{\small
\bibliographystyle{ieee_fullname}
\bibliography{egbib}
}
\clearpage

\end{document}

%% file: intro.tex
\section{Introduction}
\label{sec:intro}
Audio-driven face reenactment and talking head generation have received raising attention due to the broad killer applications in movie production, gaming, virtual digital avatars, and potentially more while we move toward the era of the metaverse. The past literature tries to advance this area from various perspectives. One line of research focuses on improving the generation quality originating from regular GAN-based methods~\cite{zhou2019talking,zhou2021pose} to pre-trained StyleGAN~\cite{alghamdi2022talking} and to the most recent Neural Radiance Field (NeRF) based methods~\cite{guo2021adnerf,shen2022learning}. Orthogonal to direction, another line of works emphasizes the importance of disentangled representation from the audio~\cite{Mittal_2020_WACV,evp2021,zhou2020makelttalk} for controlled generations. \ie,~\cite{evp2021} can predict the emotion from the audio input while~\cite{zhou2020makelttalk} can decouple the speech signal into speaker identity and the phonetic content. A closely related set of works targeting more granular controlled talking head generation by providing additional input signals. For example, PC-AVS~\cite{zhou2021pose} requires a separate video to provide the pose signal in addition to the audio clip but does not support the control of expression and blink. To remedy this, GC-AVT~\cite{Liang_2022_CVPR} requires inputting additional driving video for expression in addition to pose and audio driving clip. The technical challenge behind those approaches is how to faithfully transfer the desired driving signal (\ie, pose, expression, audio) into the results without affecting each other through intrinsic disentanglement. Although encouraging results have been reported, we argue such a setting is not practical for generalized applications. It is quite challenging, or at least labor-intensive, for novice users to find the ``best" individual driving sources for each controlled dimension to make the final talking head video overall look not only life-like but also coherent from semantic and emotional perspectives. Therefore, it is both more practical and generalizable if the setting only requires an audio signal as the driving source and expects a data-driven model to sample reasonable other facial motions irrelevant to lips.

By no means, we are the first to advocate an audio-only driving setup for talking head generation. Audio2head~\cite{wang2021audio2head} predicts the poses from audio input with an LSTM network.~\cite{lu2021live} employs an autoregressive model by assuming the poses are jointly determined by the audio and past head motions along the sequence, although the results are encouraging, the model can't generalize due to their person-specific training strategy. FACIAL~\cite{zhang2021facial} could infer the blink, but requires a reference video input rather than a one-shot reference image. There are other related works along this direction, but they either only infer one facial attribute (\ie, emotion in EVP\cite{evp2021}, pose from~\cite{wang2021audio2head}) using ad-hoc methods, or they simply do not support inferring other facial motions~\cite{wav2lip}, or treat it as a block-box mapping model~\cite{alghamdi2022talking} without explicitly respecting the one-to-many mapping nature between audio and other visual facial attributes (pose, expression, blink). A more principled solution is desired to consolidate this line of work.

In this paper, we introduce a novel framework that can holistically infer all the non-lip-related facial attributes from the audio input while maintaining accurate synchronization between audio and the corresponding visual lip motions. This is achieved by two important learning steps in our pipeline. (1) A pre-trained identity-irrelevant facial motion representation that can further be decomposed into the lip and non-lip disentangled representations. Intuitively, the lip representation should encode all the lip-related features while the non-lip one is expected to cover all the other facial motions except the lip. To promote this learning, we employ a novel orthogonal loss on top of the modified facial reenactment framework~\cite{Burkov_2020_CVPR}. (2) A novel audio-to-visual diffusion prior model is introduced to address the probabilistic sampling from audio representations to the above-learned non-lip representation. This prior is expected to solve the one-to-many mapping and provide diverse results during the inference stage. The entire pipeline can be easily built up on top of the existing framework, such as PC-AVS\cite{zhou2021pose} without heavily retraining every component. 
To sum up, we make the following contributions:
\begin{itemize}[leftmargin=*]
\item To our best knowledge, we are the first to holistically predict non-lip facial motions based on audio input only, providing good usability for reenactment or dubbing applications without extra driving video sources. Our method addresses the intrinsic one-to-many challenge in a probabilistic way, allowing diverse and realistic facial motion generation under the same audio input, as shown in Fig.~\ref{fig:teaser}. 
\item We leverage the pre-trained visual identity-irrelevant facial motion representations. And further, disentangle them into lip-related and lip-irrelevant representations through a novel orthogonal loss on top of PC-AVS\cite{zhou2021pose}. A powerful diffusion prior model is then introduced to effectively infer all lip-irrelevant facial motions for a given audio segment in the representation space.
\item We systematically evaluate the naturalness and diversity of the results with new metrics, which paves a way for future studies. Meanwhile, results show that our method can produce natural-looking head poses and facial motions without hurting the audio-lip synchronization.
\end{itemize}

%% file: relatedwork-v2.tex
\section{Related Works}
\label{sec:relatedworks}

\paragraph{Audio-Visual Cross-modal Learning} Cross-modal representation learning is a long-standing research topic, ranging from speech enhancement~\cite{gabbay2017visual}, speech source separation~\cite{lee2021looking} to synchronization~\cite{chung2019perfect,kadandale2022vocalist} and other speech disentanglement~\cite{Nagrani20d,evp2021,Mittal_2020_WACV}. Among them, EVP~\cite{evp2021} used cross-modal supervision to disentangle speech content and emotion from the audio signal with landmark as the intermediate representation. Recently, CMC~\cite{CMC2020} discussed how multi-view ``modality" can be  jointly unitized to boost the intrinsic representation through contrastive learning rather than predictive (or reconstruction) learning, it also demonstrated that the more views, the better. Concurrently, MMV~\cite{MMV2020} introduced different modality embedding graphs for effective cross-modal representations, again through contrastive learning. More recently, HCMoCo~\cite{hcmoco2022} extended similar ideas with a hierarchical strategy to learn different levels of representations for human-centric perception tasks. Although impressive results were reported, it is still unclear how it performs on face analysis tasks, not mention to on synthesis tasks. Our work shares a similar spirit but a different purpose, where we first pre-train a non-identity visual representation which is then decoupled into lip-related and lip-irrelevant representations, the latter further serves as the upper-bound learning target for subsequent audio-to-visual diffusion prior.

%Our work also follows this direction, but go one step further and introduce an iterative order-aware cross-modal contrastive learning framework, to fully respect the specificity of each modality as well as the tailored downstream reenactment-oriented tasks.

%Similarly, EVP \cite{evp2021} uses cross-modal supervision to disentangle speech content and emotion from the audio signal with landmark as the intermediate representation. Compared with EVP, the essential difference is that we jointly use visual encoding, landmark encoding as well as audio encoding to do multistage contrastive learning to learn more generalized visual and audio representations. Nevertheless, little attention has been paid to reenactment-oriented tasks powered by cross-modal visual-audio representation learning. 
\vspace{-3mm}
\paragraph{Face Reenactment \& Talking Head Generation} Face Reenactment is designed to transfer part or full facial motion from a driving source to the target video with good ID-preserved appearance and background. It can be further divided into two categories depending on whether the driving source is from video~\cite{hong2022depth,hsu2022dual,bounareli2022finding,thies2016face2face,DeepVideoPortraits,x2face2018,Zakharov_2019_ICCV,huang2020learning,Zeng2020,Burkov_2020_CVPR} or audio~\cite{IJCV2019RealisticGAN,thies2020nvp,APB2Face2020,evp2021}. Among them, audio-driven face reenactment generally aims to edit the mouth regions of the target video in order to match the input audio while leaving other facial attributes mostly unchanged, \ie, pose. EVP~\cite{evp2021} tries to infer the non-rigid facial expression in addition to lip motion from the audio input. A closely related line of work is the audio-driven talking-head generations~\cite{wav2lip,zhou2020makelttalk,zhang2021facial,alghamdi2022talking} where only one target reference face is given, hence, other face attributes including pose, expression, blink and gaze have to be either explicitly given~\cite{zhou2021pose,Liang_2022_CVPR} or partly inferred through statistical methods~\cite{zhou2020makelttalk,min2022styletalker,zhang2021facial,lu2021live,alghamdi2022talking,thies2020nvp}. Specifically, Lu \etal \cite{lu2021live} employed an auto-regressive model while Min \etal~\cite{min2022styletalker} leveraged normalized flow prior to predicting a natural-looking pose sequence from the input audio, both showed encouraging results. Compared with them, we aim to infer more diverse facial motions including pose, expression, and even blink and gaze in a holistic manner through an audio-to-visual diffusion prior model. 

%Different from all those prior works which target solving specific downstream tasks individually, we propose to pretrain both visual and audio representations holistically in one unified framework to benefit all the related face reenactment and talking head generation tasks. 
\vspace{-3mm}
\paragraph{Diffusion Generative Models}
The diffusion model~\cite{sohl2015deep, ho2020denoising, song2020denoising}, which is a likelihood-based model consisting of cascading denoising autoencoders, has recently shown great success in numerous generative tasks with different modalities including image~\cite{dhariwal2021diffusion, nichol2021glide, ramesh2022hierarchical, rombach2022high}, audio~\cite{kong2020diffwave}, video~\cite{singer2022make}, and motion~\cite{tevet2022human}. To name a few, DDPM~\cite{ho2020denoising} explored the diffusion model for unconditional image generation. GLIDE~\cite{nichol2021glide} introduced text-conditional diffusion model and showed that classifier free
guidance has better performance than CLIP~\cite{radford2021learning} guidance. DALLE-2~\cite{ramesh2022hierarchical} modified GLIDE to generate semantically consistent images conditioned on a CLIP image embedding, and proposed a diffusion prior that produces the image embedding given a text caption. MDM~\cite{tevet2022human} utilized a classifier-free diffusion-based generative model for text-to-motion and action-to-motion tasks, allowing motion completion and editing as well.

%\paragraph{}
%In this work, we decouple the lip and non-lip features from a pre-trained identity-irrelevant representation, and propose a diffusion prior to generate diverse non-lip motions from the lip feature only. To the best of our knowledge, we are the first to provide a one-to-many solution in audio-driven face reenactment with diffusion prior. Results show that our method can produce natural-looking head pose and facial motions without hurting the audio-lip synchronization.

%% file: method.tex
\begin{figure*}[h]
\centering
\includegraphics[width=0.99\textwidth,height=0.31\textwidth]{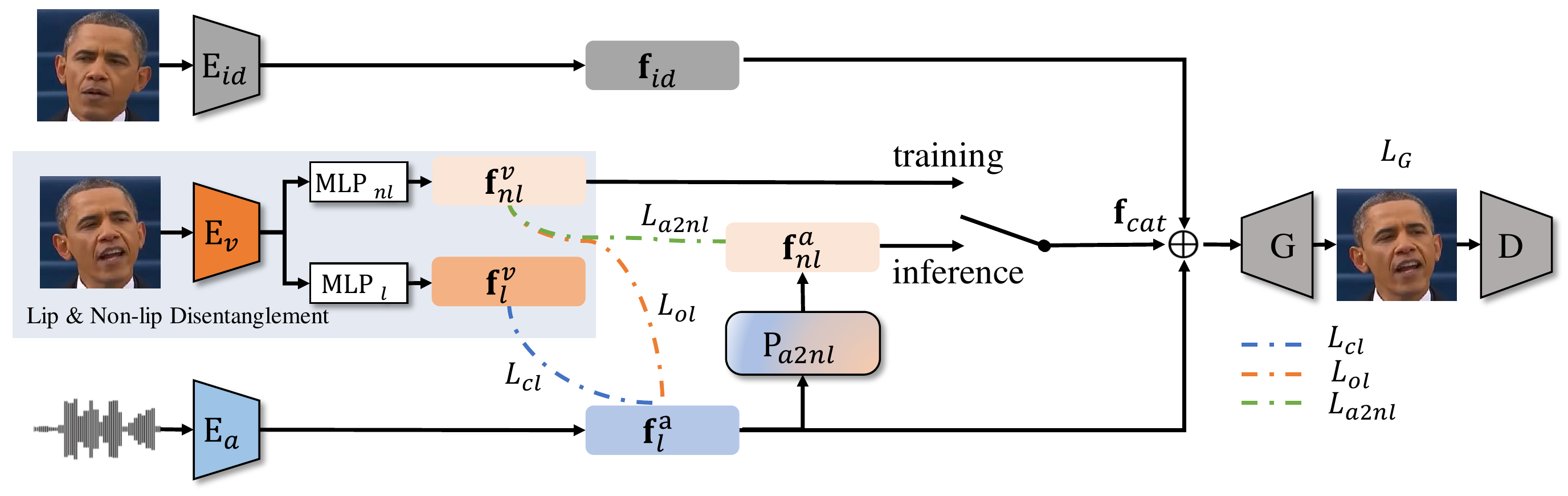}
\caption{The overall pipeline of our proposed framework. The dotted lines represent the loss functions that are used only in training, \ie, \textcolor[RGB]{68,114,196}{$L_{cl}$}, \textcolor[RGB]{237,125,49}{$L_{ol}$} and \textcolor[RGB]{112,173,71}{$L_{a2nl}$}. $L_{G}$ is the training loss for visual non-lip space. Note that there is a switch in the figure, which means the different forward processes in training and inference: in training, the concatenated feature $\textbf{f}_{cat} = cat(\textbf{f}_{id}, \textbf{f}_{nl}^{v}, \textbf{f}_{l}^a)$; at inference stage, $\textbf{f}_{cat} = cat(\textbf{f}_{id}, \textbf{f}_{nl}^a, \textbf{f}_{l}^a)$. Thus the part of Lip \& Non-lip Disentanglement is not needed anymore at the inference stage. }
% \textcolor{red}{(mark: need to optimize the structure again.)}}
\label{fig:main}
\end{figure*}

\section{Method}
Given $L$ segments of audio $A_{1:{L}}={(A_1, A_1, ..., A_{L})}$ and a reference image $I_{ref}$ as inputs, our model $\text{M}$ synthesizes video frames ${\hat{X}_{1:{L}}}=(\hat{X}_{1}, \hat{X}_{1},..., \hat{X}_{L})$ having the same identity as $I_{ref}$ and lip motion synchronized to $A_{1:{L}}$: 
\begin{equation}
    {\hat{X}_{1:{L}}} = \text{M}(A_{1:{L}}, I_{ref}).
\end{equation}

% We built our framework by modifying PC-AVS\cite{zhou2021pose}.
An overview of our proposed framework is shown in Fig.~\ref{fig:main}, which consists of three major components:
%which mainly consists of six networks: audio encoder $\text{E}_{a}$, audio2nonlip prior network $\text{P}_{a2nl}$,  visual encoder $\text{E}_{v}$, id encoder $\text{E}_{id}$, generator $\text{G}$ and discriminator $\text{D}$. Our training framework consists of three stages: 
\begin{itemize}[leftmargin=*]
    \item \textbf{Lip \& Non-lip Disentanglement} Given a pre-trained identity- and appearance-irrelevant facial motion encoder $\text{E}_{v}$, we first decompose it into two complementary features, $\textbf{f}^v_l$ for lip, and $\textbf{f}^v_{nl}$ for non-lip feature including pose, expression, blink and gaze, through $\text{MLP}_l$ and $\text{MLP}_{nl}$, respectively.
    \item \textbf{Audio to Non-lip Diffusion Prior} A diffusion prior network $\text{P}_{a2nl}$ models the one-to-many mapping from audio feature $\textbf{f}^a_l$ to the ``hallucinated" non-lip feature $\textbf{f}^a_{nl}$, which is trained to be close to its visual counterpart $\textbf{f}^v_{nl}$, allowing audio-only facial driving at inference time. 
    \item \textbf{Audio-based Talking Head Generation} Given an identity feature $\textbf{f}_{id}$, an audio feature $\textbf{f}^a_l$, and a non-lip feature $\textbf{f}^a_{nl}$ generated with the diffusion prior, we concatenate them together and feed into a GAN similar to LPD~\cite{Burkov_2020_CVPR} or PC-AVS~\cite{zhou2021pose} to output the final reenacted video.  
\end{itemize}

%For clarity, the superscript of a symbol is used to distinguish the modality it derived from and the subscript indicates whether the space is lip-related or not. Capitalize or not represent space or feature, respectively.

\subsection{Lip \& Non-lip Disentanglement}
The identity encoder $\text{E}_{id}$ is designed to capture the identity and appearance information while $\text{E}_{v}$ is instead to remove both identity and appearance while encoding all the facial motions. Both encoders are pre-trained in a similar setting as LPD~\cite{Burkov_2020_CVPR} and PC-AVS~\cite{zhou2021pose}. 
% The pretraining is followed by a training scheme consisting of audio-visual contrastive learning for lip, decorrelation of lip and non-lip features, as well as reconstruction of facial details. Losses will be described in detail as follows:
After pretraining, we conduct lip \& non-lip disentanglement consisting of audio-visual contrastive learning for lip feature learning and decorrelation for non-lip feature learning, as well as reconstruction of facial details. Details will be described as follows:
\vspace{-3mm}
\paragraph{Audio-Visual Pretraining for Audio Lip Space} An audio encoder $\text{E}_{a}$ is trained to provide accurate audio feature $\textbf{f}^a_l=\text{E}_{a}(A_{1:N})$ through contrastive learning~\cite{radford2021learning} with $\textbf{f}^v_l=\text{MLP }_l^v(\text{E}_{v}(X_{1:N}))$, where $N$ is the number of samples and $X_{1:N}$ are $N$ frames in the same video corresponding to audio $A_{1:N}$. Note that $\textbf{f}^a_l$is also called audio lip feature since it mainly contains lip related information. The contrastive loss is defined as $L_{con}(m,n) =-\frac{1}{N}\sum_{i=1}^{N}{\text{log}\frac{\exp(m_i \cdot n_i)}{\sum_{j=1}^{N}\exp(m_i \cdot n_j)}}$, resulting in a total contrastive loss between $\textbf{f}^a_l$ and $\textbf{f}^v_l$, % \paragraph{Audio-Visual Contrastive Loss for Lip} To learn an accurate audio-lip space for audio encoder $\text{E}_{a}$, the audio feature $\textbf{f}^a_l=\text{E}_{a}(A_{1:N})$ are aligned with visual feature $\textbf{f}^v_l=\text{MLP }_l^v(\text{E}_{v}(X_{1:N}))$ through contrastive loss ~\cite{radford2021learning}, where $N$ is the number of samples and $X_{1:N}$ are $N$ frames in the same video corresponding to audio $A_{1:N}$. 
% The function of contrastive loss is defined as, $L_{con}(m,n) =-\frac{1}{N}\sum_{i=1}^{N}{\text{log}\frac{\exp(m_i \cdot n_i)}{\sum_{j=1}^{N}\exp(m_i \cdot n_j)}}$ . The total contrastive loss is, 
\begin{equation}
    L_{cl} = \frac{1}{2}[L_{con}(\textbf{f}^a_l, \textbf{f}^v_l) + L_{con}(\textbf{f}^v_l, \textbf{f}^a_l)].
\end{equation}
\paragraph{Reconstruction Learning for Visual Non-lip Space} 
% To disentangle the lip-irrelevant visual feature $\textbf{f}^v_{nl}$ from the unified facial motion feature which includes lip-relevant signal, 
% For the purpose of driving non-lip motions through audio, 
To learn a good non-lip space, we propose an orthogonal loss to penalize the correlation between $\textbf{f}_{nl}^v$ and $\textbf{f}_l^a$, which uses the well-learnt audio-lip space to disentangle lip-related motions. Because we want to drive the non-lip motions through audio without hurting the audio-lip synchronization. 
% To disentangle the lip-irrelevant visual feature $\textbf{f}^v_{nl}$ from the unified facial motion feature which includes lip-relevant signal, we propose an orthogonal loss to penalize the correlation between $\textbf{f}_{nl}^v$ and $\textbf{f}_l^a$ , for the purpose of generating images with mouth movement driven only by audio.
% To extract the lip-irrelevant visual feature $\textbf{f}^v_{nl}$ and disentangle it from the lip-related visual feature $\textbf{f}^v_l$, we propose an orthogonal loss to penalize the correlation between $\textbf{f}_{nl}^v$ and $\textbf{f}_l^v$. Note that we actually decorrelate $\textbf{f}_{nl}^v$ from $\textbf{f}_l^a$ instead of $\textbf{f}_l^v$, \textcolor{red}{since $\textbf{f}_l^a$ can be obtained directly from the pre-trained audio encoder $\text{E}_{a}$. (The reason here is unreasonable.)} %The reason here is unreasonable.%
The orthogonal loss is used together with reconstruction, since there should also be a completeness constraint to avoid mode collapse of lip-irrelevant feature.
In practice, we maintain two memory banks ($\textbf{MB}$) for storing $\textbf{f}_{nl}^v$ and $\textbf{f}_{l}^a$ in previous $K-1$ steps, in order to have more samples than $N$. 
We denote the feature dimension of $\textbf{f}_{l}^a$ as $d_{a}$, and $\textbf{f}_{nl}^v$ as ${d_v}$.
The orthogonal loss is defined as follows:

% Note that our orthogonal loss should be trained with reconstruction loss together. To construct a large batch while optimizing orthogonal loss, we maintain a memory bank $\textbf{MB}$ storing the features $\textbf{f}_{nl}^v$ and $\textbf{f}_{l}^a$ in previous $K-1$ steps, where $K$ is a hyper-parameter for tradeoff between sampling accuracy and memory consumption. We denote the feature dimension of $\textbf{f}_{l}^a$ as $d_{a}$, and $\textbf{f}_{nl}^v$ as ${d_v}$, respectively. The orthogonal loss is defined as follows:

\begin{equation}
    L_{ol} = \frac{1}{d_{v}} \sum_{i=1}^{d_{v}}\sum_{j=1}^{d_{a}}P_{cor}(\textbf{f}_{nl}^v, \textbf{f}_l^a)_{(i,j)}^2,
\end{equation}
where $P_{cor}(\textbf{f}_{nl}^v, \textbf{f}_{l}^a)$ computes the Pearson correlation coefficient between $\textbf{f}_{nl}^v$ and $\textbf{f}_{l}^a$. 
We follow the reconstruction loss $L_\text{GAN}$, $L_{L1}$ and $L_\text{VGG}$ used in~\cite{zhou2021pose}. The generator $\text{G}$ receives the concatenation $\textbf{f}_{cat} = cat(\textbf{f}_I, \textbf{f}_{nl}^v, \textbf{f}_{l}^a)$ as inputs and generates final images $I_{gen}$ with modulated convolution. The discriminator ${\text{D}}$ is trained jointly by generative adversarial learning~\cite{karras2019style,goodfellow2020generative}. Note that we omit the batch-mean operation here for convenience, but in practice, all losses used here are calculated with the average in a batch. Different from~\cite{zhou2021pose}, we propose an additional gaze loss $L_{gaze}$ by utilizing a pre-trained gaze encoder~\cite{abdelrahman2022l2cs}, which calculates the $L1$ distance between the gaze features of generated images and ground-truth (GT) images $I$ as follows:
\begin{equation}
    % L_{gaze} = \left\|\Phi(I(i)) - \Phi(\text{G}(f_{cat(i)}))\right\|_1.
    L_{gaze} = \left\|\Phi(I) - \Phi(I_{gen})\right\|_1.
\end{equation}

The total loss $L_G$ is given as:

\begin{equation}
\begin{aligned}
    L_{G}   = &~\lambda_{ol}\cdot L_{ol} + \lambda_{gaze}\cdot L_{gaze} + \lambda_{L1}\cdot L_{L1} \\
              &+\lambda_{\text{GAN}}\cdot L_{\text{GAN}} + \lambda_{\text{VGG}}\cdot L_{\text{VGG}},
\end{aligned}
\end{equation}
where $\lambda$ s are weights for losses.

\subsection{Audio2nonlip Diffusion Prior}
\label{section:diffusion}
\begin{figure}[!t]
\flushleft
\includegraphics[width=0.55\textwidth,height=0.3\textwidth]{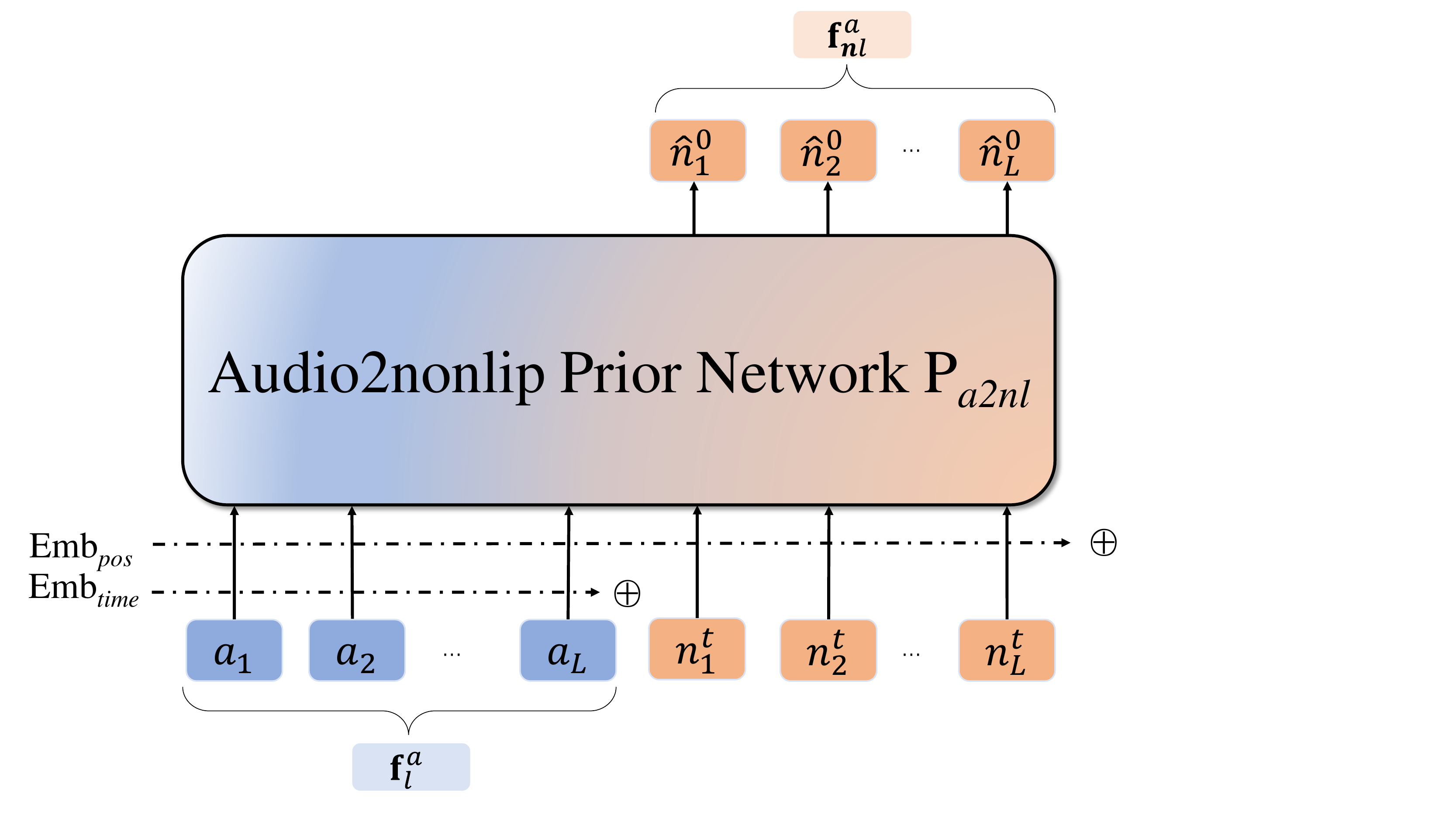}
\caption{Our \textbf{audio2nonlip} diffusion prior network $\text{P}_{a2nl}$.
% \textcolor{red}{(mark: need to optimize the structure again.)}
}
\label{fig:prior}
\end{figure}

Audio to facial motions is a one-to-many mapping problem due to the fact that there are many reasonable facial motions that can be corresponded to the same audio input. As inspired by DALLE-2~\cite{ramesh2022hierarchical} and MDM~\cite{tevet2022human}, we design an audio2nonlip diffusion prior network $\text{P}_{a2nl}$ to address this problem. Our proposed network is depicted in Fig.~\ref{fig:prior}. 
\vspace{-3mm}
\paragraph{Diffusion Model} Diffusion model is designed based on the stochastic diffusion process. The forward diffusion process is defined below:
\begin{equation}
\centering
q(n_{1:L}^t|n_{1:L}^{t-1})=\mathcal {N}(n_{1:L}^{t-1};\sqrt{{\alpha^t}}n_{1:L}^{0},(1 - {\alpha^t}){I}),
\label{equation_q}
\end{equation}
In our context, $n_{1:L}=(n_1, n_2, ..., n_L)$ is a sequence of non-lip feature $\textbf{f}_{nl}^v$. $\alpha^t$ is a hyper-parameter and $t\sim{[1, T]}$ is the time step of the diffusion process. Eq.~\ref{equation_q} can be approximated to $n_{1:L}^t \sim{\mathcal{N}(0,1)}$ if $\alpha^t$ is small enough.

During the reversed diffusion process, $\text{P}_{a2nl}$ models the audio2nonlip distribution as $p(n_{1:L}^0|a_{1:L})$, where $a_{1:L} = (a_1, a_2, ..., a_L)$ is a sequence of audio feature $\textbf{f}_{l}^a$. Instead of predicting the noise as formulated by vanilla DDPM~\cite{song2020denoising}, $\text{P}_{a2nl}$ learns to predict the initial signal itself with the simplified objective function~\cite{song2020denoising} described as follows: 
\begin{equation}
L_{simple} = \mathbb{E}_{{n^0}\sim{q({n^0}|{\textbf{f}_{l}^a})},t\sim{[1,T]}}[\left\|{n^0}-{\text{P}_{a2nl}}({n^t},t,{\textbf{f}_{l}^a})\right\|_{2}],
\end{equation}
where we use ${n^0}$ to represent $n_{1:L}^0$ for convenience. 

We implement $\text{P}_{a2nl}$ with a transformer encoder~\cite{vaswani2017attention} architecture. As shown in Fig.~\ref{fig:prior}, an audio feature $a_{1:L}$ is added to the time embedding $\text{Emb}_{time}$. Then, it's concatenated with noisy non-lip feature $n_{1:L}^t$ and added with positional embeddings $\text{Emb}_{pos}$~\cite{vaswani2017attention}. The model encodes the embeddings with bidirectional self-attention and feed-forward layers then output denoised non-lip feature $n_{1:L}^0$. 

\vspace{-3mm}
\paragraph{Velocity Loss} Note that our designed $\text{P}_{a2nl}$ denoises $n_{1:L}^t$ in a non-autoregressive way. To encourage the naturalness and coherence of generated non-lip motion, we borrow the velocity loss from MDM~\cite{tevet2022human} defined as follows:
\begin{equation}
    L_{vel} = \frac{1}{L-1}\sum_{i=2}^{L}\left\|(\hat{n}_{i}^0 - \hat{n}_{i-1}^0) - ({n}_{i}^0 - {n}_{i-1}^0)\right\|_2,
\end{equation}
where ${n}_{i}^0$ is the GT non-lip feature and $\hat{n}_{i}^0$ is the predicted denoised non-lip feature at the $i$-th position respectively. The intuition here is that the difference between adjacent non-lip features should be close to the difference in the GT. 
The total loss of this stage is defined as:
\begin{equation}
    L_{a2nl} = L_{simple} + L_{vel}.
\end{equation}

\vspace{-3mm}
\paragraph{Classifier-free Guidance} We use classifier-free guidance~\cite{nichol2021glide} for conditioned diffusion generation. In training time, we randomly set the condition to $\emptyset$ for 10\% of the samples so that $\text{P}_{a2nl}(n^t, t, \emptyset)$ approximates $n^0$. At inference stage, the output of the $\text{P}_{a2nl}$ is extrapolated further in the direction of ${\text{P}_{a2nl}(n^t, t,{\textbf{f}_{l}^a})}$ and away from ${\text{P}_{a2nl}(n^t, t,\emptyset)}$:
\begin{equation}
    \text{P}_{a2nl}(n^t, t,{\textbf{f}_{l}^a})=s\cdot{\text{P}_{a2nl}(n^t, t,{\textbf{f}_{l}^a})} + (1-s)\cdot{\text{P}_{a2nl}(n^t, t,\emptyset)},
\end{equation}
where $s$ is a scaling parameter while increasing it improves sample quality at the cost of diversity. 

\vspace{-3mm}
\paragraph{Sequential Mask Editing} To deliver a smooth non-lip facial sequence with arbitrary length, we design a mechanism to ensure continuity between generated segments, and provide an editing method to do so. Specifically, we randomly mask 90\% of the non-lip tokens $n^0$ and concatenate them with noisy non-lip tokens $n^t$ as the input of $\text{P}_{a2nl}$ at training stage. Practically, we found it easy for $\text{P}_{a2nl}$ to quickly learn how to fill in the masked region using hints from the non-lip features in the unmasked regions, generating continuous feature prediction without any extra design in training loss. As a result, it is possible to use $\text{P}_{a2nl}$ for audio-guided non-lip motion generation as well as non-lip motion editing at inference time, providing non-lip facial motion with good naturalism and diversity. 

\subsection{Audio-based Talking Head Generation}
The main difference at inference time is that non-lip visual feature $\textbf{f}_{nl}^v$ is replaced with $\textbf{f}_{nl}^a$ generated by $\text{P}_{a2nl}$ through reversed diffusion process. Thanks to the random noise introduced in the diffusion process, $\text{P}_{a2nl}$ can generate various visual non-lip features given the same audio input, resulting in diverse and reasonable reenactment videos without any extra needs of driving video sources.

To generate a video sequence longer than the maximal length of the model input with smooth transitions, the first input non-lip feature to $\text{P}_{a2nl}$ is set to the last generated non-lip feature from $\text{P}_{a2nl}$ in the previous step, to ensure continuity in non-lip facial motion.  

%% file: experiment.tex
\section{Experiments}

%\subsection{Implementation Details}
Our proposed method was evaluated from two aspects:
1) synchronization between audio and lip motion; 2) naturalness and richness of the lip-irrelevant facial motion. Both quantitative and qualitative experiments were conducted to showcase the superiority of our method. In addition, we also come up with a novel metric to measure the overall quality of generated facial motions.
\begin{table}[]\small
\centering
\setlength{\tabcolsep}{1.0mm}{
\begin{tabular}{l|ccc|ccc}
\toprule
\multicolumn{1}{c}{\multirow{2}{*}{Method}} & \multicolumn{3}{c}{VoxCeleb2}                & \multicolumn{3}{c}{VoxCeleb1}                \\
\cline{2-7}
% \multicolumn{1}{c}{}                        & FID\downarrow & Sync$_{conf}\uparrow$ & Sync$_{dist}\downarrow$  & FID\downarrow & Sync$_{conf}\uparrow$ & Sync$_{dist}\downarrow$  \\
\multicolumn{1}{c}{}                        & $\text{FID}\downarrow$ & $\text{S}_c\uparrow$ & $\text{S}_d\downarrow$  & $\text{FID}\downarrow$ & $\text{S}_c\uparrow$ & $\text{S}_d\downarrow$  \\
\midrule
GT     &  -    &   7.35   &    7.36        &    -      &   7.66   &     7.12   \\
\midrule
Wav2Lip~\cite{wav2lip}    &   22.29  &      \textbf{9.23} &  \textbf{6.33} &    44.73   &  \textbf{8.80}  & \textbf{6.75}   \\
MakeItTalk~\cite{zhou2020makelttalk}            &   19.47  &          2.03   &    12.40       &   40.24      &   2.16   &  12.34 \\
PC-AVS~\cite{zhou2021pose}    &   \textbf{14.36}  &     7.86     &    6.87      &      \textbf{35.39}    &   8.42 &       6.68    \\
Audio2head~\cite{wang2021audio2head}   &   101.37  &    6.42    &    8.26     &    103.95   &   6.65 &   8.00    \\
EAMM~\cite{ji2022eamm}     &  26.06     &   4.75     &    9.81       &      42.58     &   2.78 &      11.89  \\
\midrule
Ours + AR prior      &  24.31  &  \underline{7.23}   & 7.75 &     49.67  &      7.05    &    7.85    \\
Ours + Diff. prior    &  14.58   &     7.57      &  \underline{7.46} &  36.83   &   \underline{7.43}   &  \underline{7.58}       \\
\bottomrule
\end{tabular}
}
\caption{The quantitative results of synchronization and image quality under self-reenactment scenario. \textbf{Bold} means the best. \cite{yao2022dfa} argues that the closer to the SyncNet score of GT is actually the better, and the nearest is already highlighted with \underline{underline}.}
\label{table:main}
\end{table}
\begin{table*}[th!]\small
\centering
\begin{tabular}{l|cccc|cccc}
\toprule
\multicolumn{1}{c}{\multirow{2}{*}{Method}} & \multicolumn{4}{c}{VoxCeleb2}  & \multicolumn{4}{c}{VoxCeleb1}        \\
\cline{2-9}
\multicolumn{1}{c}{} & $\text{Var} \rightarrow$ &  
$\text{FID}_{fm}\downarrow$ & 
$\text{FID}_{\Delta{fm}}\downarrow$ &
$\text{SND}\downarrow$ & 
$\text{Var}\rightarrow$ &  
$\text{FID}_{fm}\downarrow$ & 
$\text{FID}_{\Delta{fm}}\downarrow$ & 
$\text{SND}\downarrow$ 
\\
\midrule
  GT &   1.98  &  -  & - & -  &   1.88  &  - & - &-    \\
  \midrule
  MakeItTalk~\cite{zhou2020makelttalk} &    0.67 & 4.70 & 1.74 & 6.44  & 0.80 & 4.27 & 1.07 & 5.34   \\
  Audio2head~\cite{wang2021audio2head} &  0.89  & 5.94 & 1.30 &   7.24   & 0.76 & 5.03 & 1.01 & 6.04   \\
\midrule
  Ours + AR prior &   3.07  & 5.43 & 2.22 &   7.65 & 2.92 & 6.21 & 1.68 &  7.89  \\
  Ours + Diff. prior &  1.66  & \textbf{3.81} & \textbf{1.13} & \textbf{4.94} & \textbf{1.73} & \textbf{4.17} & \textbf{0.84} & \textbf{5.01} \\
     
    \qquad  w/o $L_{vel}$  & \textbf{2.09}  & 4.02 & 1.27 & 5.29  & 2.28 &  4.65 & 1.11 & 5.76 \\
   \qquad training w/o editing &   3.09  & 6.76 & 2.00 & 8.76 & 3.06 & 6.96 & 1.70 & 8.66  \\
  \bottomrule 
\end{tabular}
\caption{\text{The quantitative results of variance and naturalness under self-reenactment scenario.} "$\rightarrow$" indicates a closer score to GT is better. }
\label{table:diversity}
\end{table*}
\vspace{-3mm}
\paragraph{Datasets} Our model was trained on VoxCeleb2 and evaluated on both VoxCeleb1 and VoxCeleb2. Here are the details of the two datasets,
\begin{itemize}[leftmargin=*, itemsep=0pt]
\item \textbf{VoxCeleb1~\cite{nagrani2017voxceleb}:} The dataset consists of 100,000 utterances from 1,251 celebrities. 100 videos with 25 identities in total were randomly chosen for the test.
\item \textbf{VoxCeleb2~\cite{chung2018voxceleb2}:} The dataset contains 1 million utterances with 6,112 identities. 500 test videos with 25 identities were randomly chosen for the test. 
\end{itemize}

\vspace{-3mm}
\paragraph{Baselines}
The following methods were selected to compare synchronization and image quality: Wav2Lip~\cite{wav2lip}, MakeItTalk~\cite{zhou2020makelttalk}, PC-AVS~\cite{zhou2021pose}, Audio2head~\cite{wang2021audio2head} and EAMM~\cite{ji2022eamm}. Notably, since MakeItTalk and Audio2head can be driven by audio only, we compare natualness and richness with them. Others were not selected due to the fact that either pre-trained models are not released or additional driving sources are required.

\vspace{-3mm}
\paragraph{Implementation Details}
Backbones of our models including $\text{E}_{id}$, $\text{E}_{v}$, $\text{E}_{a}$, $\text{G}$, and $\text{D}$ are borrowed from~\cite{zhou2021pose}. For the diffusion prior network $\text{P}_{a2nl}$, we use a 8-layer transformer~\cite{vaswani2017attention} with 512-d tokens and 1024-d fully feed-forward layers. Note that our models were trained on VoxCeleb2 only but tested on both VoxCeleb1 and VoxCeleb2.  All models were trained on 4 NVIDIA A100 GPUs. 
%\textcolor{red}{Moved to the Supplementary Materials later.}
% Two MLPs are connected to $\text{E}_{v}$ to encode disentangled lip and non-lip features and discarded with $\text{E}_{v}$ at test time.
\subsection{Quantitative Evaluation}
% \begin{figure*}[h]
% \centering
% \includegraphics[width=1\textwidth,height=0.8662\textwidth]{figures/gallery1_v7.pdf}
% \caption{\textbf{Qualitative results of our method compared to other baselines.} Each row shows nine uniformly sampled frames from videos. Here we use two audio sources to drive different identities respectively. Our method shows accurate lip-audio synchronization with diverse and natural poses and expressions. }
% \label{fig:gallery1}
% \end{figure*}

\paragraph{Evaluation Metrics}
We evaluate the performance of generated talking head from the following aspects: image quality, audio-lip sync accuracy, the variation and naturalness of non-lip facial motion. Frechet Inception Distance (\textbf{FID}) score~\cite{heusel2017gans} is used for the evaluation of image quality. A lower \textbf{FID} score indicates a lower distance between the distribution of generated images and real images. Following prior works~\cite{wav2lip,zhou2021pose}, we use SyncNet Error-Confidence ($\textbf{S}_{c}$) and SyncNet Error-Distance ($\textbf{S}_{d}$) as the indicators of audio-lip synchronization, where greater confidence or lower distance indicates better synchronization.

To evaluate facial motions such as pose, expression, and blink, we utilize a pre-trained 3D morphable face model~\cite{deng2019accurate} and include shape irrelevant 3DMM parameters
to calculate the following metrics.
\begin{itemize}[leftmargin=*, itemsep=0pt]
\item \textbf{Var}: The variance of generated facial motions, \ie, the variance of the 3DMM coefficients for each video is calculated and then averaged over the test set. A closer \textbf{Var} to GT indicates a better match to the variation of real data.
\item  $\textbf{FID}_{fm}$: FID score of 3DMM coefficients calculated as follows:
% $\textbf{FID}_{fm}(m,n)=\frac{1}{N_a}\sum_{i=1}^{N_a}({\left\|\mu_{m,i}-\mu_{n,i}\right\|_2}  + Tr(\Sigma_{m,i}+\Sigma_{n,i}-2(\Sigma_{m,i}\Sigma_{n,i})^{0.5}))$ $\mu$ is the mean and $\Sigma$ is the covariance matrix. $m$ and $n$ are the two samples to compare, with $i$ as the i-th video.
$\textbf{FID}_{fm}=\frac{1}{K}\sum_{i=1}^{K}\textbf{FID}(\beta_i)$, where $\beta_i$ is a sequence of 3DMM coefficients in the $i$-th video.
\item $\textbf{FID}_{\Delta{fm}}$: FID score of 3DMM coefficient difference between consecutive frames, \ie, \textbf{$\textbf{FID}_{\Delta{fm}}$}, which is similar to $\textbf{FID}_{fm}$ but the 3DMM coefficient difference between consecutive frames is measured, taking temporal naturalness into consideration.
\item \textbf{SND}: Our new proposed metric, denoted as \textbf{Sequence Naturalness Distance}. It is the sum of $\textbf{FID}_{fm}$ and $\textbf{FID}_{\Delta{fm}}$ indicating the difference of distribution between generated motion and GT  motion, from both spatial and temporal perspectives, \ie, $\textbf{SND}=\textbf{FID}_{fm} + \textbf{FID}_{\Delta{fm}}$. The lower \textbf{SND}, the better naturalness.
\end{itemize}
\begin{figure*}[h]
\centering
\includegraphics[width=0.9\textwidth]{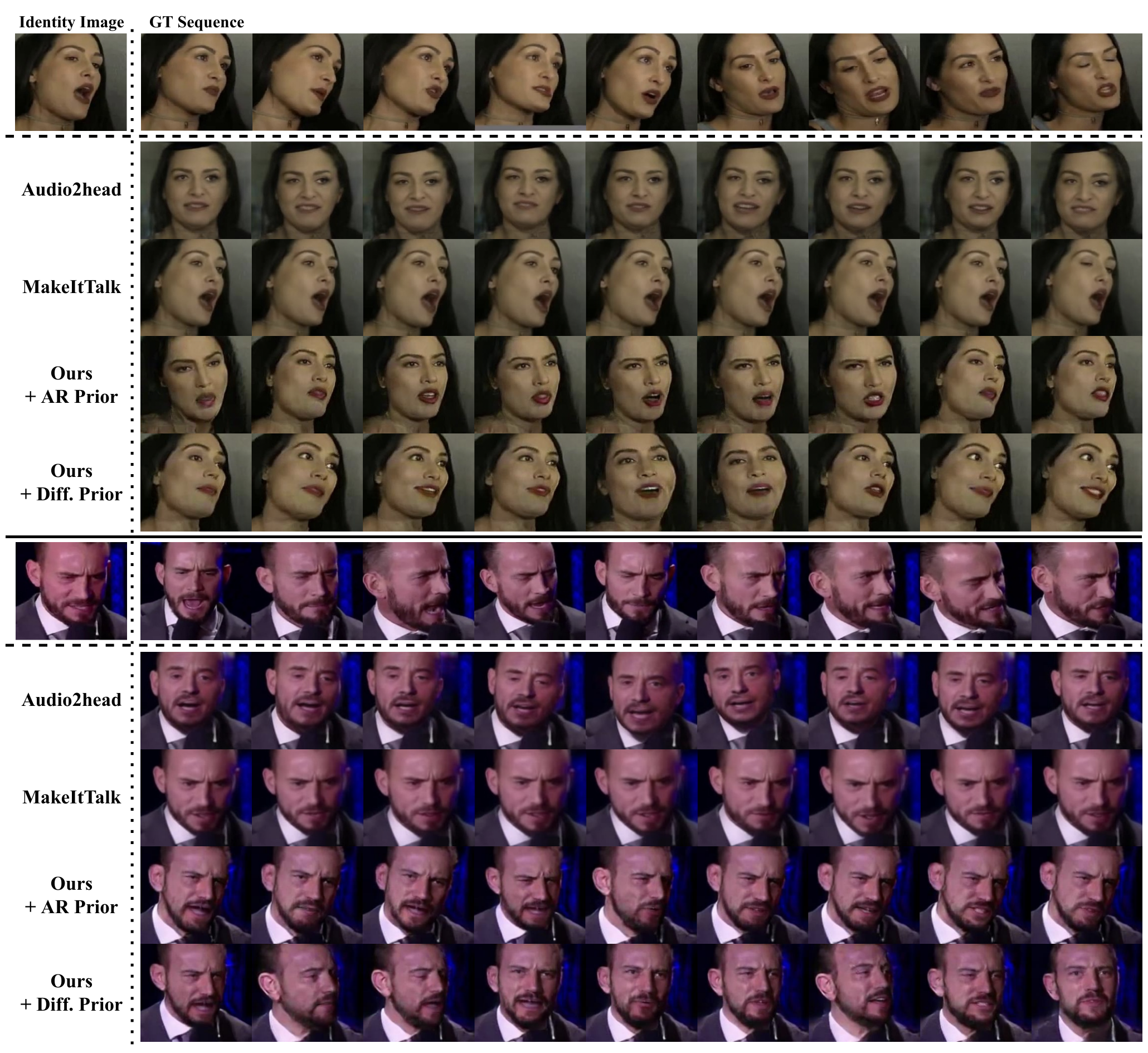}
\caption{Qualitative results of our method compared to other baselines. Each row shows nine uniformly sampled frames from videos. Here we use two audio sources to drive different identities respectively. Our method shows accurate lip-audio synchronization with diverse and natural poses and expressions. }
\label{fig:gallery1}
\end{figure*}

\vspace{-3mm}
\paragraph{Evaluation Results}

Our main results consist of two parts: audio-lip synchronization and image quality as shown in Table~\ref{table:main}; richness and sequence naturalness which are shown in Table~\ref{table:diversity}. 
% Note that Wav2Lip only predicts lower-half faces without expression and pose, and PC-AVS and EAMM use pose sequences of GT as input. Thus we only evaluate the naturalness of MakeItTalk and Audio2head in Table~\ref{table:diversity}. 

It shows that our method with diffusion prior archives relatively good synchronization scores compared to other methods in Table~\ref{table:main}. Note that Wav2Lip~\cite{wav2lip} and PC-AVS~\cite{zhou2021pose} use either additional datasets or SyncNet during training. Thus they achieve the highest SyncNet scores. However, our SyncNet scores are almost closest to the GT, indicating the most substantial synchronization ability~\cite{yao2022dfa}, which is demonstrated in Sec.~\ref{sec:Qualitative}. Meanwhile, the image quality of our model with diffusion prior is comparable to PC-AVS, better than other baselines. Since it is not our main focus to improve image quality, we build our generation loss based on PC-AVS.
% Minor score fluctuations are acceptable since we have richer expressions and movements.

In Table~\ref{table:diversity}, our method shows significantly higher variance than the other two audio-driven methods, and it is closer to  the variance of real data. Rather than slight head movement, our method can produce diverse head movements and expressions with audio only. It's notable that auto-regressive prior, trained with causal attention and regression loss, has a higher variance than GT because sometimes it generates motions with extreme pose or expression, showing a low naturalness score in Table~\ref{table:human}. For naturalness, our method achieves the lowest $\textbf{FID}_{fm}$, $\textbf{FID}_{{\Delta}{fm}}$ as well as \textbf{SND} on both VoxCeleb1 and VoxCeleb2, surpassing other state-of-the-art by a large margin. In summary,  our method can generate rich facial motions with excellent realism (\ie, a lower distribution distance) and naturalness.

To quantify the one-to-many diversity, we also investigated \textbf{Multimodality} from MDM~\cite{tevet2022human} to measure the average distance of 3DMM parameters between different runs given the same inputs. However, due to the lack of one-to-many data, \ie, multiple videos corresponding to the same audio, we only calculate it as 2.31 for future comparison. 
% Note that although auto-regressive prior shows high variance, its extreme pose and expression lead to a bad SND score, same as naturalness score shown in Table~\ref{table:human}. Thus SND is a reliable metric for audio-driven methods to evaluate naturalness
%As for the \textbf{Multimodality} metric, due to the fact that other methods are unable to generate diverse motions given the same audio as input, we only calculate our own number. Our \textbf{Multimodality} score of diffusion prior on VoxCeleb2 is 2.31. To sum up, our method archives the best naturalness with relatively good diversity, synchronization ability, and image quality. 
% not only in the rationality of temporal and spatial distribution of generated motions, but also in the diversity. 
% To sum up, our method archives the best diversity with relatively good synchronization ability and image quality. 

% \begin{figure*}[h]
% \centering
% \includegraphics[width=1\textwidth,height=0.8662\textwidth]{figures/gallery1_v7.pdf}
% \caption{\textbf{Qualitative results of our method compared to other baselines.} Each row shows nine uniformly sampled frames from videos. Here we use two audio sources to drive different identities respectively. Our method shows accurate lip-audio synchronization with diverse and natural poses and expressions. }
% \label{fig:gallery1}
% \end{figure*}

\subsection{Qualitative Evaluation}\label{sec:Qualitative}
Two audio-driven video sequences of our method verses other baselines are shown in Fig.~\ref{fig:gallery1}. It's clearly shown that our method is capable of producing much more natural-looking and diverse facial motions than other baselines, including pose, expression, blink and gaze, while maintaining accurate audio-lip synchronization compared to the GT.
\vspace{-3mm}
\paragraph{User Study}
\begin{table}[t!]\small
\centering
\setlength{\tabcolsep}{1.5mm}{
\begin{tabular}{l|ccc}
\toprule
Method & Sync Accuracy & Naturalness & Richness \\
\midrule
GT & 4.61 & 4.41 & 4.18 \\
\midrule
MakeItTalk~\cite{zhou2020makelttalk} & 1.97 & 2.35 & 2.10 \\

PC-AVS~\cite{zhou2021pose}* & 3.13 & 2.68 & 2.68 \\

Audio2head~\cite{wang2021audio2head} & 2.56 & 2.35 & 2.34   \\
\midrule
Ours + AR prior& 3.46 & 2.53 & 3.37 \\
Ours + Diff. prior& \textbf{4.08} & \textbf{3.82} & \textbf{3.68} \\
\bottomrule
\end{tabular}
}
\caption{Human evaluation on generated samples on VoxCeleb2. * means that PC-AVS was evaluated on cross-reenactment scenario.}
\label{table:human}
\end{table}
We invited 20 subjects for human evaluations, focusing on three aspects: 1) the accuracy of audio-lip synchronization; 2) the naturalness of non-lip motions and coherency between non-lip motions and audios; 3) the richness of non-lip motions. 50 videos are generated from audio and identity in the test set using the following methods: our diffusion prior, our auto-regressive prior, MakeItTalk~\cite{zhou2020makelttalk}, Audio2head~\cite{wang2021audio2head} and PC-AVS~\cite{zhou2021pose} (driven by poses from another video). After shuffling the generated videos, each annotator rates from 1 (bad) to 5 (good) according to Mean Opinion Scores (MOS) rating protocol. Table~\ref{table:human} tells that our model with diffusion prior archives the best synchronization, naturalness and richness among all the models. Our model with auto-regressive prior archives relatively good richness compared to the remaining models, we attribute it to the exposure bias problem~\cite{bengio2015scheduled} brought by auto-regressive generation, which leads to high richness but low naturalness. Note that PC-AVS shows incompatible result in Table~\ref{table:human}  as compared to Table~\ref{table:main} on synchronization scores, indicating that choosing pose driving signals from a random video may lead to synthesis results with bad audio-lip synchronization.

It's interesting that our proposed \textbf{SND} scores reflect the \textbf{naturalness} of user study to some degree. Nevertheless, we leave it to future work for rigorous study of their correlation through more human evaluations.

\subsection{Ablation}
\begin{table}[]
\centering
\setlength{\tabcolsep}{1.5mm}{
\begin{tabular}{l|cc|cccc}
\hline
\multicolumn{1}{c}{\multirow{2}{*}{Method}} & \multicolumn{2}{c}{Lip Only}  & \multicolumn{4}{c}{Non-lip Only}        \\
\cline{2-7}
% \multicolumn{1}{c}{} & $\text{Sync}_{conf}\uparrow$ & $\text{Sync}_{dist}\downarrow$ & $\text{Dis}_{blink}\downarrow$ & $\text{Dis}_{gaze}\downarrow$ & $\text{Dis}_{exp}\downarrow$ & $\text{Dis}_{pose}\downarrow$  \\
\multicolumn{1}{c}{} & $\text{S}_{c}\uparrow$ & $\text{S}_{d}\downarrow$ & $\text{B}_{d}\downarrow$ & $\text{G}_{d}\downarrow$ & $\text{E}_{d}\downarrow$ & $\text{P}_{d}\downarrow$  \\
\hline
GT & 7.35  & 7.36 & - & - & - & -  \\
\hline
w/o $L_{ol}$ & 2.84 & 13.02 & 0.0074 & 0.1960 & 0.0568 & 0.0021  \\
w/o $\textbf{MB}$ & \textbf{7.69} & \textbf{7.33} & 0.0112 & 0.0315 & 0.0875 & 0.0293   \\
Ours & \underline{7.48} & \underline{7.54} & \textbf{0.0067} & \textbf{0.0163} & \textbf{0.0558} &  \textbf{0.0019} \\
\hline
\end{tabular}
}
\caption{Evaluation of disentanglement between lip and non-lip on VoxCeleb2. $\text{B}_{d}$ and $\text{G}_{d}$ denote L2 distances of 2D landmark~\cite{yu2021heatmap}  of eyes and iris, respectively, while $\text{E}_{d}$ and $\text{P}_{d}$ denote L2 distances of 3DMM coefficients of poses and expressions accordingly. Note that $\textbf{f}^v_{nl}$ is fixed while tested on lip only metrics, and $\textbf{f}^a_l$ is fixed while tested on non-lip only metrics for fair comparisons.}
\label{table:disentangle}
\end{table}
Table~\ref{table:disentangle} discusses the performance of lip \& non-lip disentanglement. While tested on lip only metrics without $L_{ol}$, SyncNet scores get lower, which indicates that $\text{G}$ tends to generate mouth movements with information extracted from visual modality instead of audio. Thus, lip \& non-lip space are not fully disentangled. Without \textbf{MB}, the performance of driving non-lip motions gets worse. Because it is hard to use a large batch to compute correlation, resulting in a not well decoupled non-lip space.

\begin{figure}[h]
\centering
\includegraphics[page=1,width=0.45\textwidth,height=0.2104\textwidth]{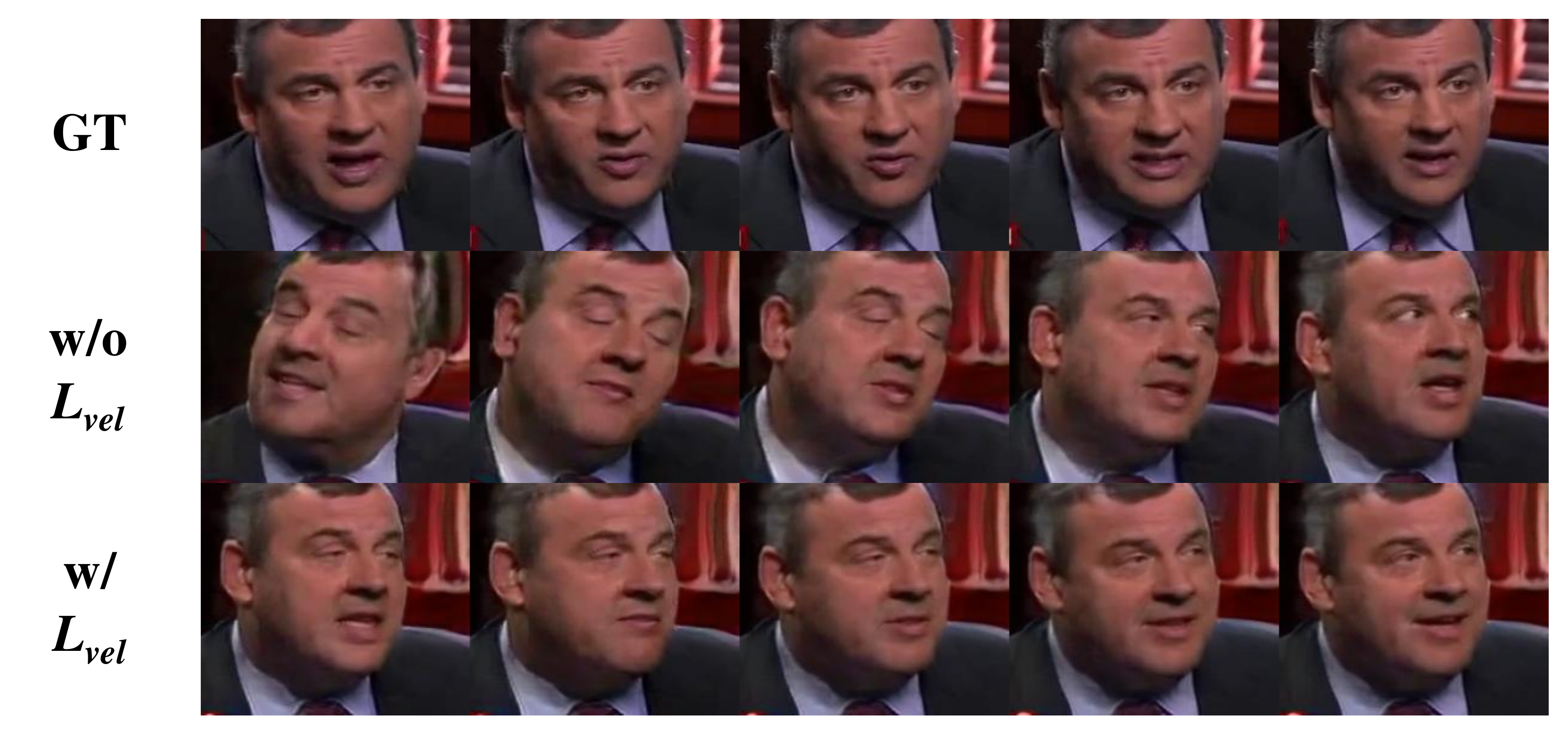}
\caption{Ablation study of our diffusion prior $\text{P}_{a2nl}$ w or w/o $L_{vel}$. Each row shows five uniformly sampled frames from videos. } 
\label{fig:ablation_vel}
\end{figure}

\begin{figure}[h]
\centering
\includegraphics[page=2,width=0.45\textwidth,height=0.2104\textwidth]{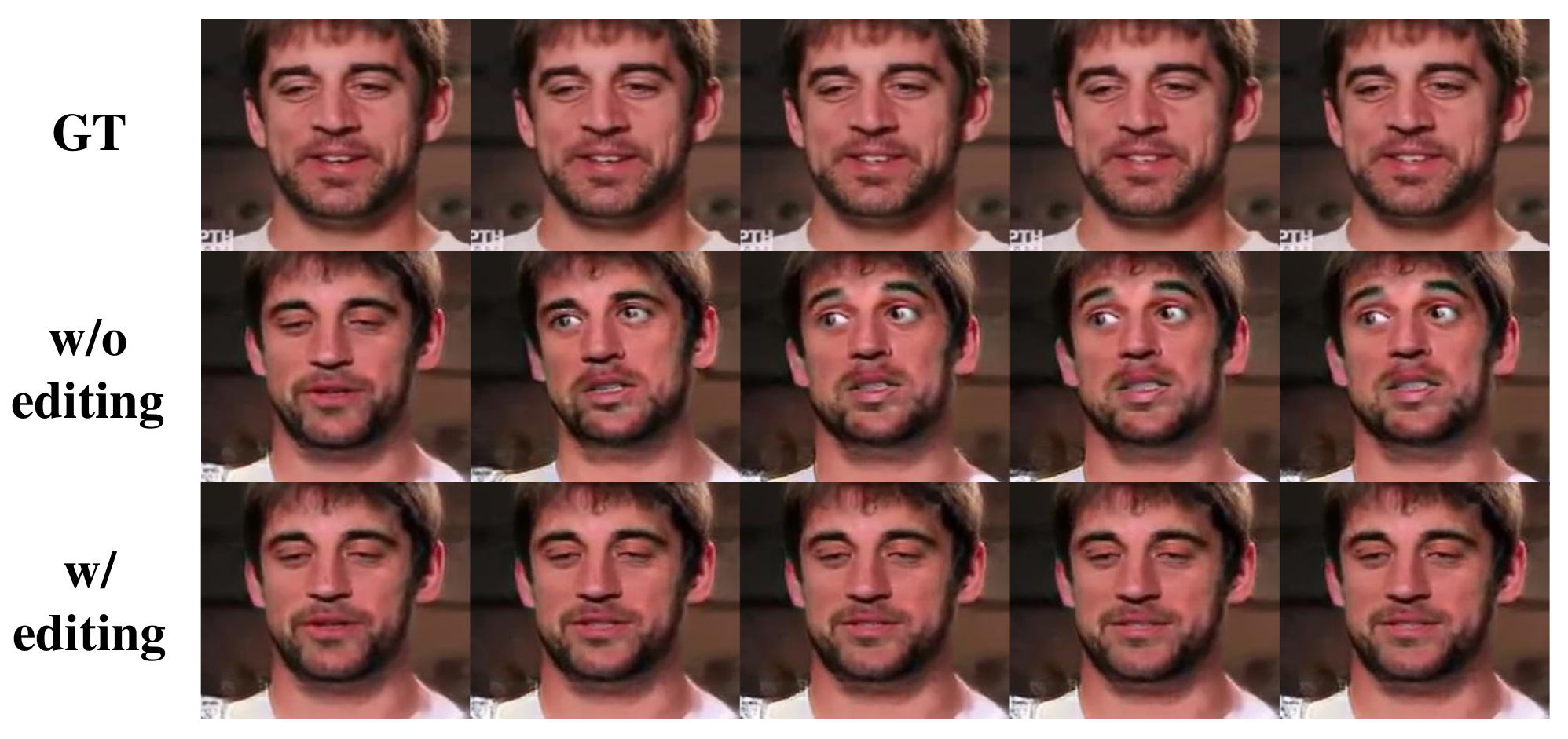}
\caption{Ablation study of our diffusion prior $\text{P}_{a2nl}$ trained w or w/o editing. Each row shows five adjacent frames. }
% There seems to be a frameskip between the first and second frame in the $2^{nd}$ row.
\label{fig:ablation_editing}
\end{figure}

We conducted ablation study for $\text{P}_{a2nl}$ from three aspects: 1) auto-regressive prior versus diffusion prior; 2) with/without velocity loss $L_{vel}$; 3) training with/without editing mechanism. Diffusion prior shows siginificant improvements on naturalness scores including $\text{FID}_{fm}$, $\text{FID}_{\Delta{fm}}$ and \textbf{SND}, as compared to autoregressive prior shown in Table~\ref{table:diversity}, with slightly better synchronization ability in Table~\ref{table:main}. Fig.~\ref{fig:ablation_vel} shows that without $L_{vel}$, the prediction of non-autogressive diffusion prior model $\text{P}_{a2nl}$ becomes unstable, which leads to a larger variance and worse \textbf{SND} scores in Table~\ref{table:diversity}. While training without the editing mechanism mentioned in Sec.~\ref{section:diffusion}, we applied editing only during sampling as in MDM~\cite{tevet2022human}, and observed that it jitters between adjacent frames as shown in Fig.~\ref{fig:ablation_editing}, resulting in a poorer \textbf{SND} score in Table~\ref{table:diversity}. Note that our model trained with editing mechanism generates smoother results.

% \begin{figure}[h]
% \flushleft
% \includegraphics[width=0.45\textwidth,height=0.355\textwidth]{figures/disentangle.png}
% \caption{\textbf{Qualitative results of Lip \& Non-lip Decoupling.} Yellow arrows are for winks, blue are for mouths, red are for gazes and green are for expression.}
% \label{fig:disentangle}
% \end{figure}